\documentclass[fleqn,10pt]{olplainarticle_arxiv}

\title{Review of Machine Learning for Micro-Electronic Design Verification}

\author[1]{Christopher Bennett}
\author[2]{Kerstin Eder}
\affil[1]{christopher.bennett@bristol.ac.uk}

\date{\today}


\DraftwatermarkOptions{stamp=false}

\begin{abstract}
Microelectronic design verification remains a critical bottleneck in device development, traditionally mitigated by expanding verification teams and computational resources.  Since the late 1990s, machine learning (ML) has been proposed to enhance verification efficiency, yet many techniques have not achieved mainstream adoption.  This review, from the perspective of verification and ML practitioners, examines the application of ML in dynamic-based techniques for functional verification of microelectronic designs, and provides a starting point for those new to this interdisciplinary field.
Historical trends, techniques, ML types, and evaluation baselines are analysed to understand why previous research has not been widely adopted in industry.  The review highlights the application of ML, the techniques used and critically discusses their limitations and successes. Although there is a wealth of promising research, real-world adoption is hindered by challenges in comparing techniques, identifying suitable applications, and the expertise required for implementation.  
This review proposes that the field can progress through the creation and use of open datasets, common benchmarks, and verification targets.  By establishing open evaluation criteria, industry can guide future research.  Parallels with ML in software verification suggest potential for collaboration.  Additionally, greater use of open-source designs and verification environments can allow more researchers from outside the hardware verification discipline to contribute to the challenge of verifying microelectronic designs.
\end{abstract}

\begin{document}

\flushbottom
\maketitle
\thispagestyle{empty}

\section{Introduction}
The production of micro-electronic devices is a multi-billion-pound industry where the cost of design errors found after tape-out is high.  As a result, sources suggest that up to 70\% of development time in a microelectronic design project is invested in verification to find bugs before production~\cite{foster2022}.  Historically, step changes in verification techniques have enabled the electronics industry to keep pace with the greater complexity of electronic designs.  For instance, using simulation-based verification to support manual inspection, use of hardware emulation to speed up simulations, introducing UVM to standardise the way verification environments are built and reusued, and using constrained random instead of expert design instruction sequences.  The EDA (Electronic Design Automation) verification industry is now asking whether Machine Learning will be the next step change.

The rising cost and development time for microprocessor verification is driven by customer demand.  Customers want devices with greater functionality, performance and lower cost.  To meet these demands, microelectronic designs are becoming increasingly complex.  
The industry is seeing a trend towards system-on-chip designs and integrating heterogeneous components with multiple IPs from different manufacturers.

This complexity is compounded by often incomplete functional specifications, leading some to remark that device specifications are becoming more of a statement of intent rather than a rigorous design reference.  Consequently, the likelihood of errors has increased at all stages of production due to misinterpretation of specifications and mistakes in design and synthesis.  This places significant pressure on verification teams to ensure correct operation amid growing design complexity and higher error rates.  The trend seen over the last decade of hiring more verification engineers and investing in costly simulation time~\cite{foster2022} is not viewed as sustainable.  As a result, many in the EDA industry look to machine learning to assist in the verification effort.

The increasing complexity of designs and rising cost of verification is not the only motivation for using machine learning.  The creation of open-source designs, such as those based on RISC-V, have enabled non-specialists to create commercial chips.  While the open-source movement fosters innovation, it also introduces risks.  These non-specialists may lack the the verification expertise and resources of the traditional manufacturers, but the cost of design errors remains high. Therefore, the proliferation of open-source designs emphasises the need for design verification techniques that are efficient, effective and accessible.  Machine learning is a tool with the potential to address these needs.  

Machine learning involves making predictions and classifications based on data. The design and verification of electronic devices generate large amounts of often labelled data, including specifications, code, and test results. This makes machine learning well-suited for microprocessor verification. Recent advances in reinforcement learning~\cite{Silver2017} for gameplay and large language models for generative AI have garnered significant attention, leading to substantial interest from the EDA industry in using machine learning to reduce the time, cost, and bottlenecks associated with verification.

Although interest in this area is growing, it is not new. For over 20 years, both academia and industry have explored incorporating machine learning into the verification process. Despite this, the verification of electronic devices still relies heavily on expert-directed random simulations. The key question is why research in this area has struggled to gain adoption in real-world projects. This review aims to address this question. Specifically, it takes the perspective of an EDA practitioner, highlighting the verification challenges where machine learning has been applied, the techniques used, and critically discussing the limitations and successes.

Unlike recent reviews of machine learning in EDA that took a broad view of the EDA process, this review focuses specifically on the use of machine learning for the functional verification of pre-silicon designs using dynamic (simulation-based) techniques.  Traditionally cited as the greatest bottleneck in microprocessor development, it is also an area where two decades of research have not translated to industrial practice.  

Written for practitioners in both industry and academia, the review supports the future application of cutting-edge ML techniques to break through from concept to industrial practice.

\subsection{Scope}
\begin{figure}[t]
    \centering
    \includegraphics[width=0.8\linewidth]{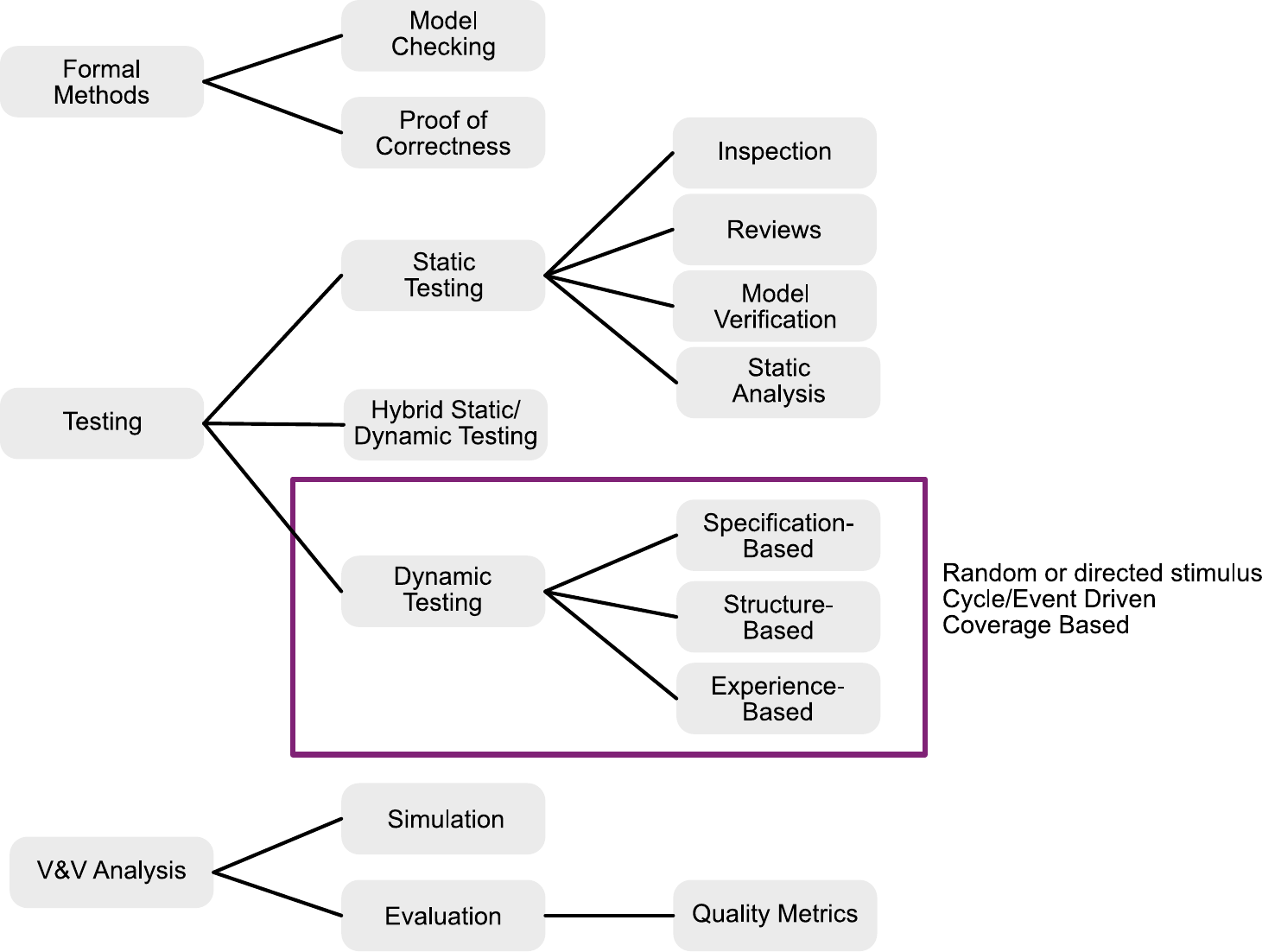}
    \caption{The role of dynamic based test methods within verification and validation.  Adapted from ISO/IEC/IEEE 29119-2:2013, ``Software and systems engineering — Software testing — Part 1: Concepts and definitions''.  The scope of the review is enclosed in the purple rectangle.}
    \label{fig:isoscope}
\end{figure}

This review focuses on how machine learning can be used in a dynamic functional verification process for microelectronic designs.  Figure~\ref{fig:isoscope} shows the range of verification activities and the scope of this review.  Dynamic verification is distinguished by the use of test-methods based on applying random or directed stimulus in cycle or event driven simulations~\cite{Molina2007}.  In the context of electronic design verification, these dynamic methods are distinct from static and formal verification methods, which instead use techniques including SAT and BDD Solvers, Theorem Proving, Property Checking, Model Checking and Formal Assertion Checking~\cite{Molina2007}.  There are also hybrid methods that use a combination of both static and dynamic techniques, however these are not included in the review. 

In the electronic design industry, dynamic verification is a process rather than a singular activity.  Authors have expressed different views of the activities that constitute this process. For example, \citep{Ismail2021b} describes the process as consisting of Stimulus \& Test Generation, RTL modelling, Coverage Collection, Assertions Checking, Scoreboarding \& Debugging. Whereas in \citep{Wu2024} the process is described differently.  Design Characterisation and Coverage Prediction are added as activities, Debugging is split into Detection, Localisation and Debug, and Assertion Checking and Scoreboarding are not included.   Different definitions of the verification process are not surprising.  Verification is a process to check the correctness of a device against its specifications. Therefore, the activities that constitute a dynamic verification process vary to reflect the needs of a specific project.

This review focuses on a critical part of the verification process that would be applicable, in whole or in part, to most projects.  Specifically, the application of stimuli and recording of coverage.  At its simplest, a design is simulated, and its output is recorded in response to various stimuli.  If the response does not match the expected behaviour, then an error is recorded.  The primary challenge for verification teams during this activity is to generate input stimuli that efficiently test a design against its specification. 

\subsubsection{Exclusions}
There has been research interest in using machine learning for hardware verification for approximately 20 years, resulting in a wide and varied literature.  Exhaustively covering this literature is impractical.  Therefore, we excluded some methods and activities in the dynamic-verification process where machine learning can be used.

The first exclusion is the use of machine learning with formal techniques, such as accelerating formal analysis and selecting the best formal technique to use.  Formal techniques are an important part of block-level verification, especially for safety-critical designs, because they can exhaustively explore the state-space of a design.  However, formal techniques do not currently scale to complex designs and are less widely used than dynamic techniques on industrial projects.  Formal techniques also draw on a different set of analytical tools, including SAT solvers, and covering these would distract from the core aims of the review.  Hybrid techniques that mix formal with dynamic techniques were not included for the same reason.

To ensure a focus on design-based verification, we excluded research related to hardware implementation.  This includes work related to the use of ML for design analysis, such as predicting the physical area occupied by a design from its RTL description \cite{Zennaro2018} and verifying layout~\cite{Francisco2020}.
Machine learning also has applications for activities that support finding errors, such as design emulation, creating test benches and creating coverage models from specifications. However, these are beyond the scope of this review.
We also excluded material relating to troubleshooting since this is the step that occurs after the detection of an error.  Troubleshooting includes the use of machine learning for triage, root-cause analysis and debug.

Finally, the review excludes using machine learning to verify non-functional specifications, including power, security and robustness to soft errors~\cite{Li2024}. For instance, using ML to create a bespoke model of power use or find patterns in RTL code indicating trojan hardware~\cite{ Yasaei2021} is excluded. 

\subsubsection{Inclusions}
Traditionally, the scope of machine learning includes supervised, unsupervised and reinforcement learning techniques.  In this review, we also chose to include the use of evolutionary algorithms in our definition of machine learning.  These algorithms are heuristic-based searches and are not always covered by a definition of machine learning.  However, the use of evolutionary techniques is common in research for dynamic-based verification, and excluding these techniques would prevent traditional machine learning being compared with the state of the art.

The scope of this review also encompasses a select number of machine learning applications that extend beyond traditional definitions of functional verification. This includes the closure of structural coverage models, such as finite state machines, and code coverage models, including branch and statement coverage. Additionally, applications of machine learning for test pattern generation using pre-silicon simulations are considered. The rationale for including these applications is that the machine learning techniques and methodologies involved are sufficiently similar to those used in functional verification, making them of interest to practitioners, even if the exact application may differ.

\subsection{Contributions}
Machine learning has a long history in the verification of electronic hardware as an academic endeavour but not in widespread industry practice. Recent developments in machine learning have further propelled academic interest in the topic.  However, there is a risk of perpetuating the status quo where developments in verification research fail to gain real-world adoption.  This review aims to mitigate the risk by contributing a platform for both academic researchers and industry to understand the state of the art, helping researchers to understand the limitations of existing approaches and industry to find the material relevant to the specific challenges they face.  It does so by taking a systematic, critical and detailed look at the research material from the perspective of an industry practitioner. 

This review builds on previous surveys covering the use of machine learning in the electronic design process.  Each of these surveys presented a different perspective, including large surveys covering the entire EDA process~\cite{Huang2021}, pre and post-silicon verification with a bias towards formal and hybrid techniques~\cite{Jayasena2024}, and the use of both static and dynamic techniques~\cite{Wu2024}.  These large surveys have had a wide scope and tended towards high-level and broad observations of the state of the art.  Supporting the large surveys are smaller surveys which focus on a single area for the use of machine learning in verification.  For instance, the use of Reinforcement Learning, Neural Networks and Binary Differential Evolution Algorithms~\citep{Vangara2021}, and the application of ML from an industry perspective~\cite{Yu2023}.   There have been relatively few large surveys that specialised in one element of the hardware verification process.  The closest are~\cite{Ioannides2012}, which does not cover the recent developments in machine learning, and \cite{Ismail2021b} which has a similar scope and includes an in-depth discussion of neural network-based test generation techniques.      

Unlike these prior works, this review takes a systematic, critical and detailed look at the use of current machine learning techniques to support simulation-based design verification, including a detailed examination of the how previous research has been evaluated.  While previous surveys have forwarded an understanding of the breadth of Machine Learning in EDA design, the specialism of this review enables greater depth and analysis, crucial to ensuring the application of new ML techniques does not experience the limitations of prior work and that developments break through into industrial practice.  

The use of Machine Learning in simulation-based verification is a large topic, and like previous surveys, this work does not claim to be exhaustive.  However, we present and follow a systematic methodology to enable others to replicate our work and build upon it by expanding the analysis to new areas for the EDA process.

The review is written to support industry practitioners or academic researchers using ML in their verification activities.  Consequently, unlike previous surveys, this review is written from the perspective of having a problem to solve and the need to understand the state of the art, limitations of techniques, and open challenges.  This top-down approach enables discussion and navigation of the topic guided by need.  It is distinct from a ``bottom-up'' approach that starts with a pool of literature and forms classifications based on them, which suits an understanding of the literature but is perhaps less helpful to a practitioner.

\noindent In summary, the contributions of this review are:
\begin{itemize}[nosep]
    \item[-] Written for industry practitioners looking to use ML in their verification activities and academics looking to understand the state of the art and open challenges.
    \item[-] A specialist review of machine learning in dynamic-based micro-electronic design verification to enable greater depth, commentary and synthesis.  
    \item[-] Written from a top-down perspective starting with the industrial development process and need.  
    \item[-] A commentary on coverage models and evaluation is included.  Both of these are crucial to assessing the success of simulation-based testing and have not been covered in previous work.
    \item[-] A clear methodology to collect prior work and a quantitative analysis to identify trends and gaps in the research.  
\end{itemize}

\subsection{Review Structure}
The review is structured as follows.  The methodology and scope of the review are given in Section~\ref{sec:methodology}.  Section~\ref{sec:background} familiarises the reader with the core concepts necessary to understand the field of dynamic-based hardware verification.  A quantitative assessment of the research material is given in Section~\ref{sec:quantitiveanalysis}, followed in Section~\ref{sec:usecases} by problems the research aims to address and the characteristics of an ``ideal'' dynamic test platform.  Coverage models and the application of machine learning to coverage closure are discussed in Section~\ref{sec:qualcoverageclosure}, and Sections~\ref{sec:bughunting}~to~\ref{sec:testsetoptimisation} discuss respectively the application of machine learning to finding bugs, detecting faults and optimising test sets. 
 Section~\ref{sec:evaluation} discusses the hardware and metrics used by research to evaluate techniques.  The review ends in Section~\ref{sec:challengesopportunities} with a summary of the open challenges and opportunities.

\section{Paper Collection and Methodology} \label{sec:methodology}

\subsection{Methodology Used to Collect Material}
The review adopts a methodology similar to that used in \cite{Fontes2023} for a survey of machine learning in software verification.  The prior art was sampled using a structured search of literature from the  IEEE~Xplore \footnote{\url{https://ieeexplore.ieee.org/Xplore/home.jsp}} and  Web of Science databases.  Results were restricted to accessible material written in english.  The format of the search string was $\mathit{problem} \times \mathit{application} \times \mathit{technology}$.  
\begin{quote}
    (("All Metadata":rtl OR "All Metadata":eda OR "All Metadata":"functional verification" OR "All Metadata":"functional coverage") 
        AND ("All Metadata":verification OR "All Metadata":validation) 
        AND ("All Metadata":"machine learning" OR "All Metadata":"reinforcement learning" OR "All Metadata":"deep learning" OR "All Metadata":"neural network" OR "All Metadata":bayesian) )
\end{quote}
Material was also sampled from the proceedings of Design and Verification Conference and Exhibition USA because it is historically well supported by industry.  Due to limitations in the search functionality, the following terms were searched independently and the results combined: \textit{coverage, machine learning, reinforcement learning, deep learning, neural network, Bayseian, genetic algorithm}. 

\begin{figure}[h]
    \centering
    \includegraphics[width=1.0\linewidth]{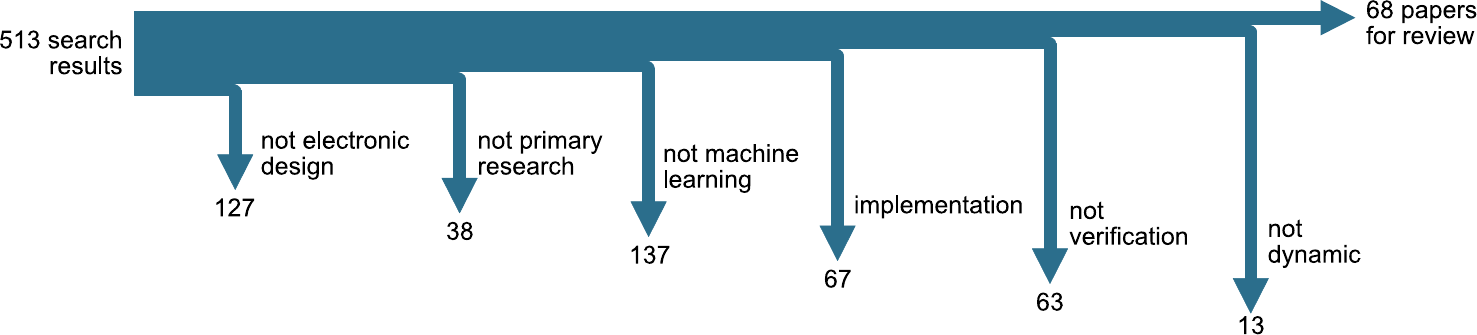}
    \caption{Methodology used to filter search results.}
    \label{fig:filteringmethodology}
\end{figure}

The search initially returned 513 results. These were filtered by first removing any that did not relate to the electronic design \emph{process}, including research that proposed hardware designs to accelerate machine learning algorithms. Next, papers were removed that were not primary research (including surveys and commentaries) or did not feature machine learning in the primary aim of the paper. Finally, work relating to physical hardware design, such as layout, routing, analogue modelling, or analogue design, or did not relate to verification were removed from the results. Decisions were based on the abstract, title, keywords and a paper's introduction.  When the classification of material was unclear, it was reviewed collaboratively between the authors.  The remaining papers were read in detail, and relevant information was tabulated, including coverage models and the type of machine learning used.
The resulting dataset (paper references, classifications and tables) is available from the corresponding author upon request and will be available for download at \url{https://data.bris.ac.uk/data/} in due course.

\subsection{Research Questions}
The research questions the review aims to answer are listed below.  These support the high level aim of reviewing the state of the art for the use of machine learning (ML) in EDA verification.  
\begin{itemize}
    \item[-]RQ1: How has ML been used to perform or enhance the dynamic-verification process for electronic designs?
    \item[-]RQ2: How is the deployment of ML evaluated?
    \item[-]RQ3: Which specific ML techniques were used to perform or enhance coverage closure?
    \item[-]RQ4: What are the limitations and open challenges in integrating ML into EDA verification?
\end{itemize}

\section{Background} \label{sec:background}
This section introduces dynamic-based verification concepts and terminology, particularly for those from a machine learning background. Experienced practitioners in microelectronic design and verification may wish to skip to Section 4.

\subsection{Verification in the Digital Design Process}
The digital design process is divided into Front-End and Back-End activities~\cite{Danciu2022}.  Front-End activities focus on what the design will do, while Back-End activities determine how it will do it.  During the Front-End stage, the design's functional behaviour is developed according to its specification and represented at different levels of abstraction.  There are three levels in common use: Register-Transfer (RTL), gate, and transistor~\cite{Parthasarathy2022a}.  Some authors also include a higher level of abstraction called behavioural representation, written in high-level languages like SystemC or C++.  The Back-End stage transforms the abstract design into a physically implementable form through activities such as floorplanning, placement, routing, and timing analysis.

Verification is a process to establish the correctness of a device against its specification throughout the design process. This review focuses on verification during the Front-End stage, we refer to this as \textbf{functional verification} to emphasise it aims to check a design's behaviour rather than its implementation.  Descriptions of the modern functional verification process are given in~\cite{Danciu2022} and~\cite{Fine2006}.  Here, a device is referred to as the Design Under Verification (DUV) to emphasise it is a design description, not a physical device. The term Design Under Test (DUT) is also used in the literature.

There are three common types of verification commonly used to establish functional correctness: dynamic, hybrid, and static verification. Dynamic verification applies stimulus to a simulation of a design and checks if the design’s output matches the specification. Static verification uses analytical methods like model checking which do not simulate the design. Static methods can exhaustively prove a design's behaviour for all inputs and states but are computationally infeasible for complex designs due to the state explosion problem. Dynamic verification, while not exhaustive, is more scalable and the most widely used method. Hybrid methods combine simulations with static analysis to balance scalability and rigorous proofs, such as using simulations of real behaviour as the starting point for proofs rather than all possible behaviour (some of which may not be realisable).

\subsection{Coverage Models and Closure}
Dynamic verification methods cannot exhaustively verify complex designs, especially within time-constrained commercial projects. Instead, verification teams use coverage models to focus efforts on design elements of interest. A coverage model defines the scope of a verification task, and it is used to measure progress. The dynamic verification process is considered complete when all elements in the coverage models are tested (covered) and the correct behaviour observed~\cite{Romero2009}, a milestone known as coverage closure.  Most ML-enhanced verification techniques reviewed use coverage models both in the learning process and for performance evaluation. Examples are discussed in Section~\ref{sec:coveragemodels}

Coverage models are divided into structural and functional types.  \textbf{Structural models} are based on the design description and examples include statement, conditional, branch, toggle, and state machine coverage. These models are generated automatically and are used to track how thoroughly the design has been executed during testing.  By comparison, \textbf{Functional models} derive from the DUV's specification and track whether the design is functionally correct. Definitions and examples of functional and structural coverage models can be found in~\cite{Piziali2007}.

Functional coverage models are usually created manually by the verification team. A verification plan, derived from a DUV’s specification, identifies features and associates them with one or more coverage models.  A typical project may have hundreds of coverage models, with some overlapping.  Therefore, a feature and its associated states can appear in multiple coverage models.  Consequently, a sequence of inputs to a DUV can cover multiple states in a coverage model and many models.  One challenge for ML-enhanced verification techniques is to operate with a range of coverage models, both structural and functional.

There are different types of functional models commonly seen in research using ML for electronic-design verification:

\begin{itemize}
    \item[-] \textbf{Cross-product coverage models}: These are named groups of states in the DUV’s state space. They define cover points, which are specific points in the design to monitor, such as the values of signals or variables. A coverage cross is the cross product of two or more cover points, and a cross-product coverage model is a collection of these crosses~\cite{Mandouh2018}.  A simplified version defines cover points in isolation, without relating them to other signals or variables.
    
    \item[-] \textbf{Assertion-based models}: An assertion expresses a property of the design, such as a safety property (something that should never happen) or a liveness property (something that should eventually happen). The purpose of an assertion model is to report the occurrence of an expected event~\cite{Piziali2007}. Assertions are broadly divided into those defined only over primary input signals and those defined over input, internal, and output signals~\cite{ Li2023}. An advantage of assertion models is their suitability for static-based techniques, making them advantageous in projects that use both formal and test-based methods. However, this review found they are rarely used with machine learning for dynamic verification, potentially due to their association with static-based techniques. They are used in hybrid methods, such as Goldmine~\cite{Liu2012}, an ML-based technique that uses simulation traces and formal methods to create assertions automatically.
\end{itemize}

Some applications may define alternative functional coverage models. For example,~\cite{Romero2009} applies ML to verify a design at a system level, and the research uses \textbf{Modular coverage}, to record when a specific block (module) is activated.

It is common to refer to the coverage of a test. In the case of functional coverage, it measures how well a test covers part of the functional specification. In the case of structural coverage, it measures how well the test covers the implementation of the specification~\cite{Simkova2015}. The coverage of a test can be viewed as the percentage of the coverage model a test covers.

Structural and functional coverage models have limitations. Structural coverage models are easy to create but only reveal how much of the design has been tested, not whether its behaviour is correct. Conversely, functional coverage models track how much of the specified behaviour has been tested but do not measure the quality and completeness of the verification environment~\cite{Yang2013}. Functional models are usually created manually, which introduces the possibility of human error and limits the scope to the behaviour defined by the verification team. Therefore, achieving coverage closure with both structural and functional models does not guarantee a bug-free design.

\textbf{Coverage closure} aims to test all reachable states within a coverage model, but \textbf{quality of coverage} is also important. Each point in a coverage model should be accessed multiple times through different trajectories originating from previous states, and the frequency of visits to each point should be evenly distributed. While most machine learning applications reviewed have tackled the issue of coverage closure, few studies address the requirements for multiplicity and distribution. Examples that do include~\cite{Fine2003b, Farkash2015}.

\subsection{Testing in Dynamic-Based Verification}
Testing is one technique in the suite of verification methods, and it is  central to dynamic-based verification.  In testing, inputs are applied to the Design Under Verification (DUV), and its responses are recorded. Typically, a test bench is used, which includes a test generator, a simulator, the DUV, an output recorder, and a golden reference model (ground truth) to check the correctness of the DUV's output. 

The primary goal of testing is to identify bugs in the design, prioritising high-priority bugs that relate to fundamental errors. Tracing the root cause of a test failure can be complex and time-consuming. Although not the focus of this review, machine learning has been used to aid debugging~\cite{Stracquadanio2024, Shen2019}. Additionally, test failures can occur due to errors in the verification environment rather than the design itself, and machine learning techniques have been employed to predict which is the source of these failures~\cite{Wahba2019}.

Dynamic-based testing is often divided into a directed and volume stages~\cite{Gogri2022}. The directed stage focuses on establishing basic functionality and targeting expected bugs. This is followed by the volume stage, which uses automatically generated tests to uncover bugs arising from rare conditions that are difficult to predict. The volume stage occupies most of the simulation time, although not necessarily human resource, and is the primary focus of machine learning approaches.

A third stage, regression testing, involves periodically running a set of tests to verify the current state of the design. Often part of a continuous integration/continuous development workflow, regression testing repeats previously completed tests to ensure that design changes have not introduced new errors~\cite{Kumar2023}. The challenge for regression testing is to select the smallest number of tests that can effectively expose any new errors. Examples of machine learning applications addressing this challenge are discussed in Section~\ref{sec:testsetoptimisation}.

In addition to the different stages of testing, there are various methodologies for creating the stimuli needed to drive the Design Under Verification (DUV). Three traditionally used approaches are expert-written tests, pseudo-random tests, and coverage-directed tests~\cite{Guzey2008}. Writing effective tests by hand requires expert knowledge and time. Therefore, the micro-electronic verification industry conducts volume testing using test generators to automatically create stimuli for the DUV. These generators are not purely random.  Instead, they incorporate domain knowledge to generate stimuli that are more likely to find errors in the DUV's design.  This knowledge is traditionally encoded by experts, although research has used ML to extract this knowledge automatically~\cite{Katz2011}.  The verification team can then parametrise these generators to target specific behaviours.  When the parameterisation is constraints, the process is known as \textbf{Constrained-Random test Generation} (CRG) or  Constrained Random Testing (CRT)~\cite{Phogtat2024}.

A central challenge in dynamic-based verification is the (sometimes) complex relationship between the inputs a DUV receives, the states it enters, and the outputs it produces.  Each time a test is simulated on a device, information is gained about this relationship that can be used to guide future testing. \textbf{Coverage-Directed test Generation} (CDG) uses constrained test generators where constraints are set based on the coverage of previous tests. These constraints can be set by experts or machine learning algorithms and are updated throughout verification to target different functionalities.

Even with a single set of constraints, the output of a constrained random test generator (and the behaviour of the DUV) can be varied by changing the random seed and initial state. Industrial test generators can have over 1000 constraints, making their configuration non-trivial. 
Machine learning can be used to parametrize constrained test generators (Section~\ref{sec:testdirection}), therefore it is important to realise the potentially large feature space and the need to identify the relevant features (parameters) to control.

Coverage-directed generation is a mature industry-standard approach, well-defined in the SystemVerilog language~\cite{Ieee1800-2023} and Universal Verification Methodology~\cite{Ieee1800-2-2020}, used by approximately 70\% of real-world projects~\cite{foster2022}. Its advantages include the ability to generate tests for devices with many inputs, cover functionality in a balanced way, and quickly create many test cases~\cite{Romero2009}. However, it is inherently computationally inefficient due to its reliance on pseudo-random generation. The effectiveness of a parameterisation to increase coverage decreases over time~\cite{Guo2010}, and the approach can be ineffective for hitting  specific coverage points (e.g., coverage holes)~\cite{Masamba2022a}. Compared to expert-written tests, Coverage-directed generated tests are often longer, less targeted, and use more simulation resources to achieve the same result~\cite{Shen2005}.  One topic of research is to use machine learning to increase the efficiency of CDG and enable tighter control over its output.

\textbf{Coverage-Directed Test Selection} is a variant of CDG where pre-existing tests are selected based on their potential to increase coverage. This approach is especially beneficial when tests are computationally cheap to generate but expensive to simulate, and it is a focus of ML research~\cite{Masamba2022a}.

\subsection{The Verification Environment}
\begin{figure}[t]
    \centering
    \includegraphics[width=0.90\linewidth]{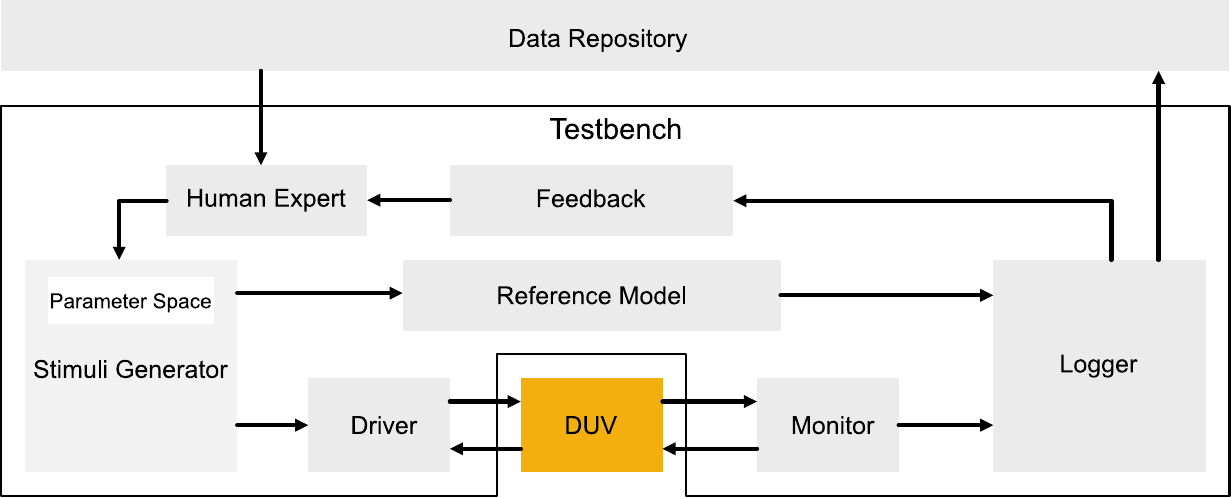}
    \caption{A conventional test bench used in the test-based verification of microelectronic designs.  The test bench is configured for Coverage-Directed Generation using a parameterised stimuli generator and where human expertise (not machine learning) is used to control the generation of stimuli to the Design Under Verification (DUV).}
    \label{fig:generictestbench}
\end{figure}

The typical dynamic-verification environment makes use of a testbench as shown in Figure~\ref{fig:generictestbench}. The stimuli source can be either expert-written instruction sequences or those generated by a constrained-random test generator. These stimuli are translated into inputs compatible with the Design Under Verification (DUV), which is then simulated, and its response is monitored. A reference model, or golden model, checks if the response aligns with the design specifications.  Most research using machine learning methods interface with a variant of this environment, discussed in Section~\ref{sec:qualcoverageclosure}.

Dynamic-based verification also uses a repository to store information necessary for replicating tests and results from previous runs. These repositories typically contain large amounts of labelled data, from which machine learning techniques can be trained to, for instance, select tests to rerun after a design change or predict whether a new test will verify a specific DUV behaviour.

Finally, a single instantiated test or set of constraints may reveal multiple instances where the DUV's input does not produce the expected output. These errors could be due to a mistake (bug) in the DUV design or an issue in the verification environment. The test-based process described here is part of a workflow where test outcomes are analysed to identify, diagnose, and correct errors in both the DUV design and the test bench.

For machine learning practitioners, it is crucial to establish what a test constitutes in this environment, as a test description can be part of the training data, model output, or both.  Depending on the author, the term ``test'' can refer to a single input, a sequence of inputs, a complete program, or a parameterisation including constraints and random seeds. A test may also involve the configuration of the DUV~\cite{Zheng2023}, and a transaction can be expressed at different levels of abstraction, from a bit-pattern to a high-level instruction. To avoid confusion, we define the following terms:
\begin{itemize}
    \item [-] \textbf{Test-template}: The parameterisation that biases a test generator, including the random seed, constraints, and any additional information needed to generate output.
    \item \textbf{Instantiated-test}: A sequence of inputs created by a test generator to be applied to a DUV.
    \item \textbf{Constraints}: The parameterisation applied to a constrained random test generator.
    \item \textbf{Directed Test}: A test program written by an expert, denoting a sequence of inputs to a DUV.
    \item \textbf{Transaction}: An instruction or command expressed at a high level of abstraction.
    \item \textbf{Stimuli}: A low-level, bit-pattern, input to the DUV.
\end{itemize}

\subsection{The Challenge of Coverage-Directed Verification}
The primary challenge for test-based, dynamic verification is to find all bugs in a design using the least amount of human and computational resources.  Ultimately, this is what most research using machine learning for functional verification aims to achieve (Section~\ref{sec:benefitsofml}).  

In coverage-directed verification, progress is often tracked by the cumulative percentage of coverage points hit vs the number of simulations performed. The goal is to shift the curve to the left, achieving higher coverage in fewer simulation cycles. Alternatively, a more granular view of coverage is to associate each ``test'' with the coverage points it hits.   We found examples of machine learning techniques using each view of coverage as part of reward, fitness or cost functions, or as labels for supervised techniques.  

An alternative view based on the number of points covered per test cycle is proposed in~\cite{Farkash2015}.  This view reveals waves where each peak is the covering of a new area of functionality.  Different test scenarios can be fingerprinted by these waves.  There were no examples found by this review of research into alternative views of coverage and their impact on learning, suggesting it is an under explored area.

Hitting the last 10 percent of coverage points is often more difficult because these represent rare corner cases. Some research concentrates specifically on hitting the remaining coverage-holes after a high percentage of coverage has been achieved~\cite{ Masamba2022b}.  Authors refer to the redundancy rate as the proportion of instantiated-test inputs that do not increase coverage~\cite{Guo2010}. The redundancy rate usually increases as verification progresses, indicating that the efficiency of computational resources decreases when hitting the hard-to-reach coverage points.

\section{The Distribution of Research by Topic}\label{sec:quantitiveanalysis}
\begin{figure}[t]
    \centering
    \includegraphics[width=0.8\linewidth]{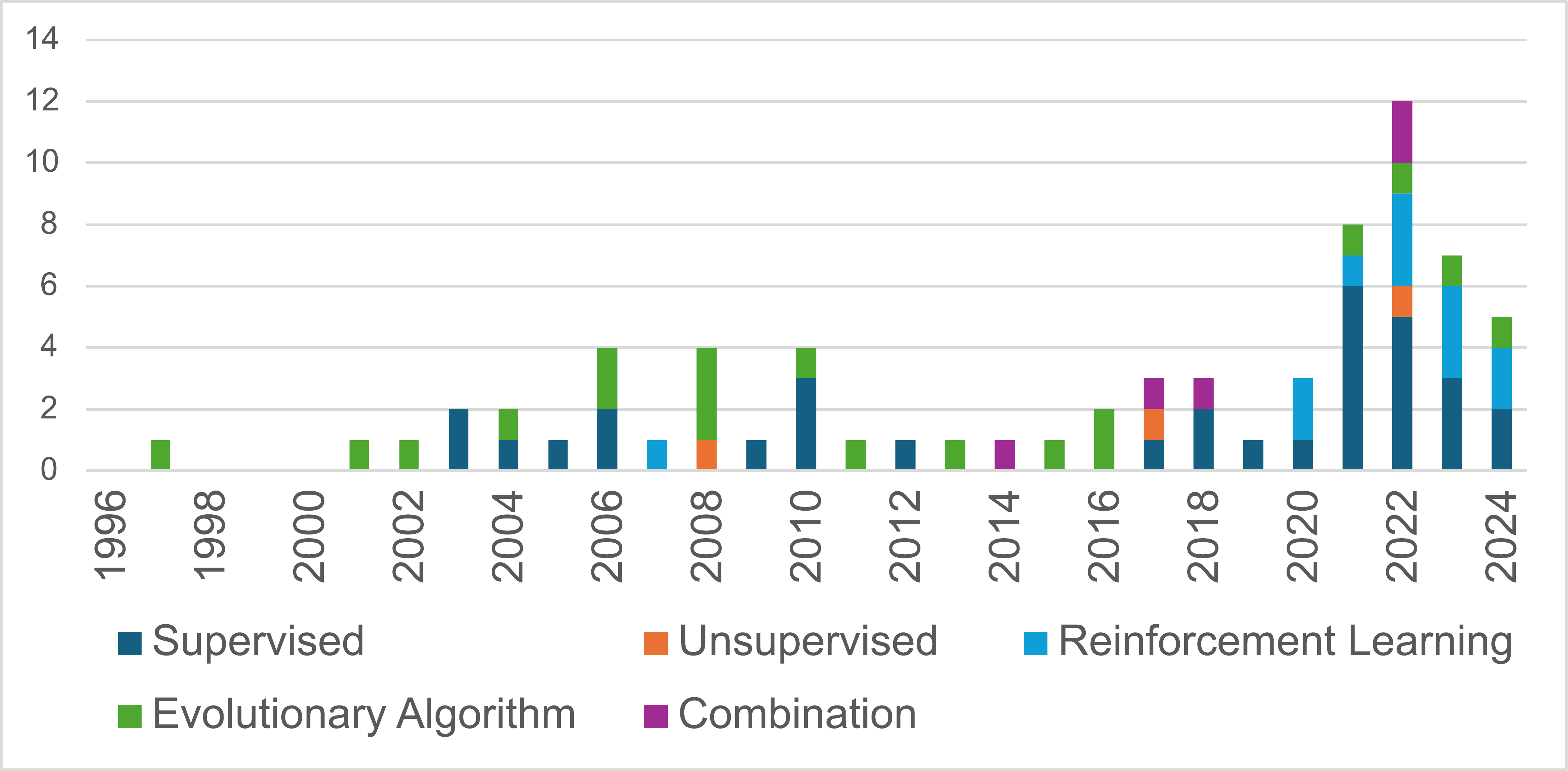}
    \caption{Number of papers by year and machine learning type.}
    \label{fig:mltypetimeline}
\end{figure}

The methodology outlined in Section~\ref{sec:methodology} produced a sample of the literature.  In this section, we analyse this sample and make observations related to quantitative measures of the material to highlight trends and gaps.
The earliest work found that applied machine learning to EDA verification was the use of evolutionary algorithms~\cite{Smith1997} in 1997. From 2001 to 2020, a steady interest in the topic is seen, and evolutionary algorithms are the most frequently used technique.  In 2018, a shift occurred where research switched to using supervised techniques.  Despite the work in 2007, it was not until 2020 that the use of reinforcement learning (RL) was seen.  A step change is seen in 2021 where the number of papers is more than double that seen in any previous year, and this increased interest has been sustained to 2024 (Figure~\ref{fig:mltypetimeline}).
 
In the work surveyed, the authors did not propose machine learning techniques specifically for EDA verification. Instead, adaptations of techniques developed in other fields were used. Therefore, these trends reflect interest in and use of machine learning more broadly. Reinforcement learning and unsupervised techniques are potentially under-represented in the sampled research. However, the wide availability of labelled data and the extra expertise to set up RL explains why supervised techniques are prevalent in recent research efforts.

\begin{figure}[h]
    \centering
    \includegraphics[width=0.6\linewidth]{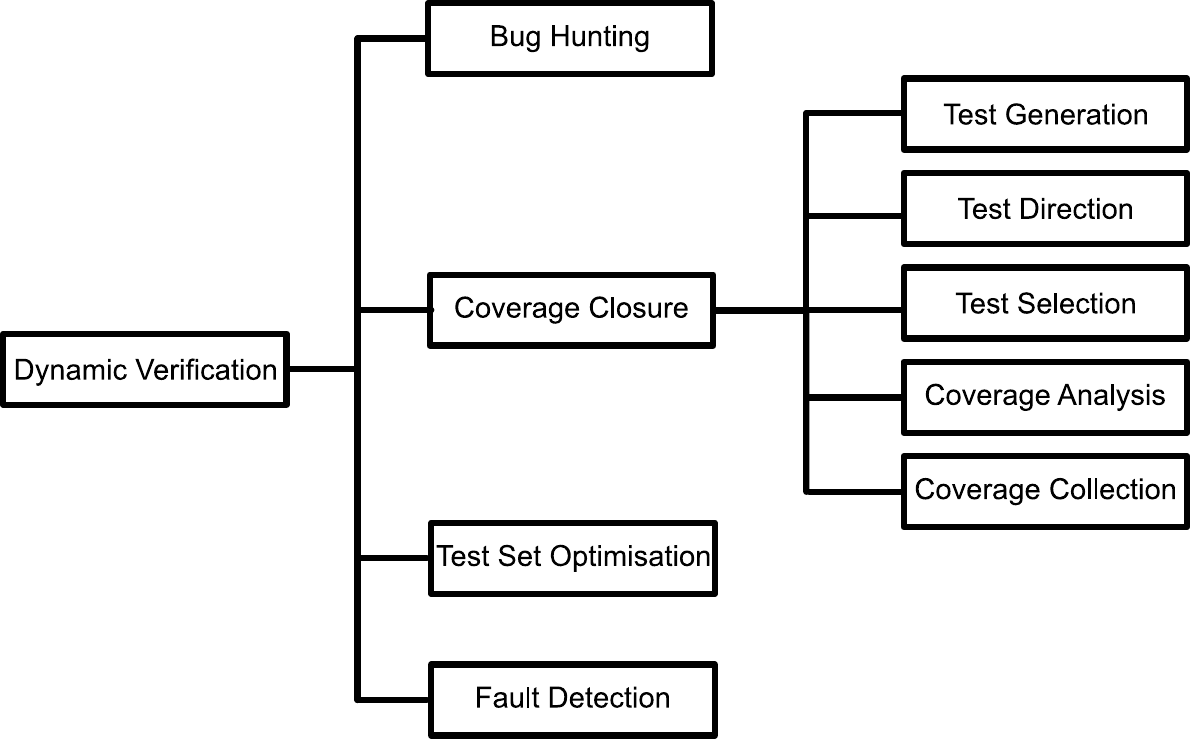}
    \caption{Verification activities using machine learning within the sampled research material for the functional verification of digital designs using dynamic-based methods.}
    \label{fig:mlusetaxonomy}
\end{figure}

In the sampled literature, we found examples of four real-world dynamic-verification activities supported by machine learning techniques. These were bug hunting, coverage closure, test set optimisation and fault detection (Figure~\ref{fig:mlusetaxonomy}).  In bug hunting, a verification engineer seeks to predict or uncover new bugs based on prior experience of where these bugs may occur.  Coverage closure also uncovers bugs, but its aim is different.  Coverage closure measures verification progress against pre-defined metrics.  With respect to the terminology used in software testing~\cite{ieee291192013}, bug hunting can be viewed as similar to experience-based testing and coverage closure as requirements-based testing.  Fault detection aims to create inputs to a design that will trigger bugs.  Unlike coverage closure and bug hunting, the bugs in fault detection are pre-defined and the inputs are primarily intended for later use.  For example, to test post-silicone designs or field testing.  Coverage closure, but also bug hunting and fault detection can create a large number of tests.  Test set optimisation is the activity of testing the same design behaviours but with less simulations.  Test set optimisation is synonymous with regression testing, an industry practice where previously completed tests are re-run to verify design changes.

\begin{figure}
    \centering
    \begin{subfigure}[b]{0.47\linewidth}
        \centering
        \includegraphics[width=\linewidth]{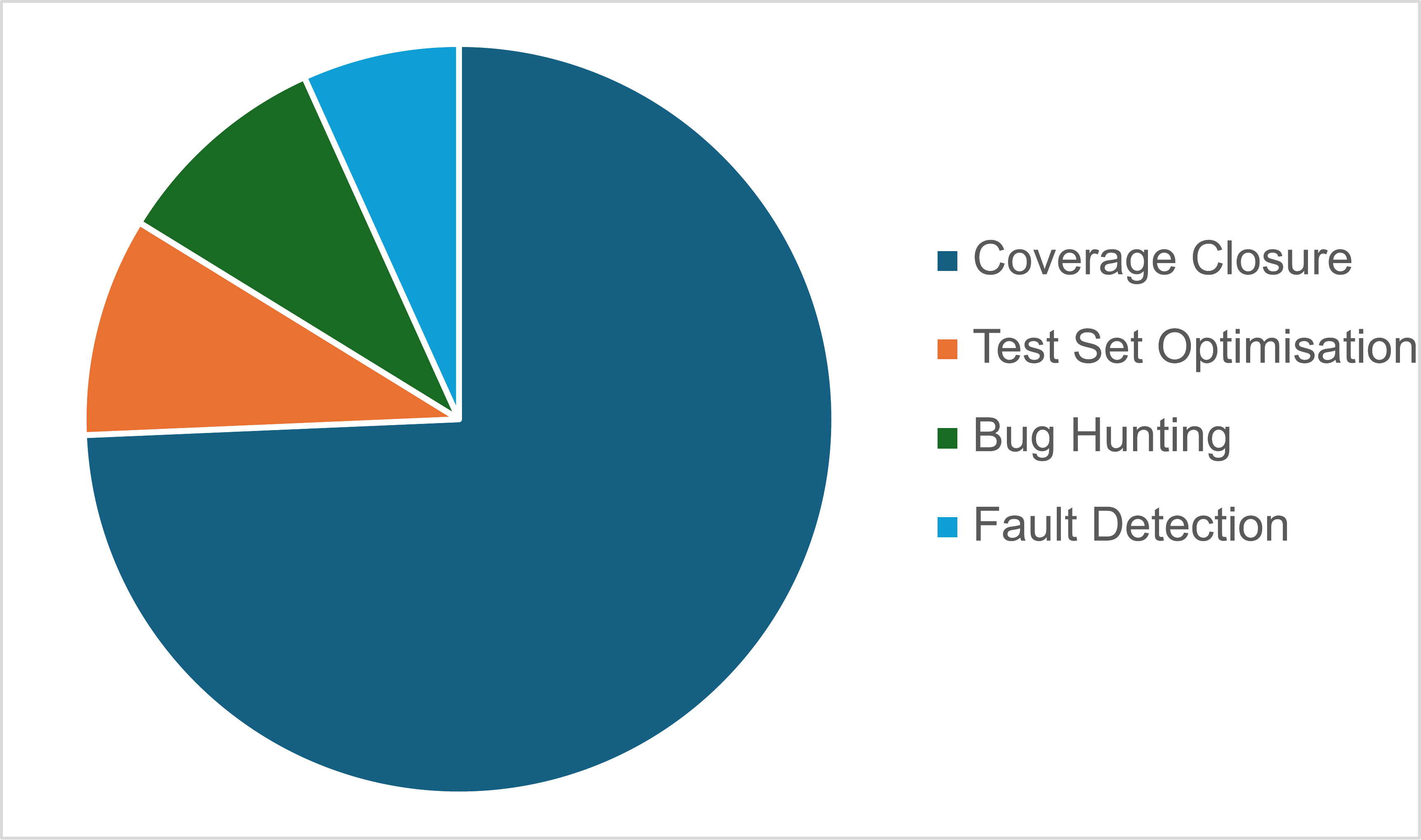}
        \caption{Proportion of papers by activity within verification. \phantom{xxxxx}}
        \label{fig:verif_process_pie}
    \end{subfigure}
    \hfill
    \begin{subfigure}[b]{0.47\linewidth}
        \centering
        \includegraphics[width=\linewidth]{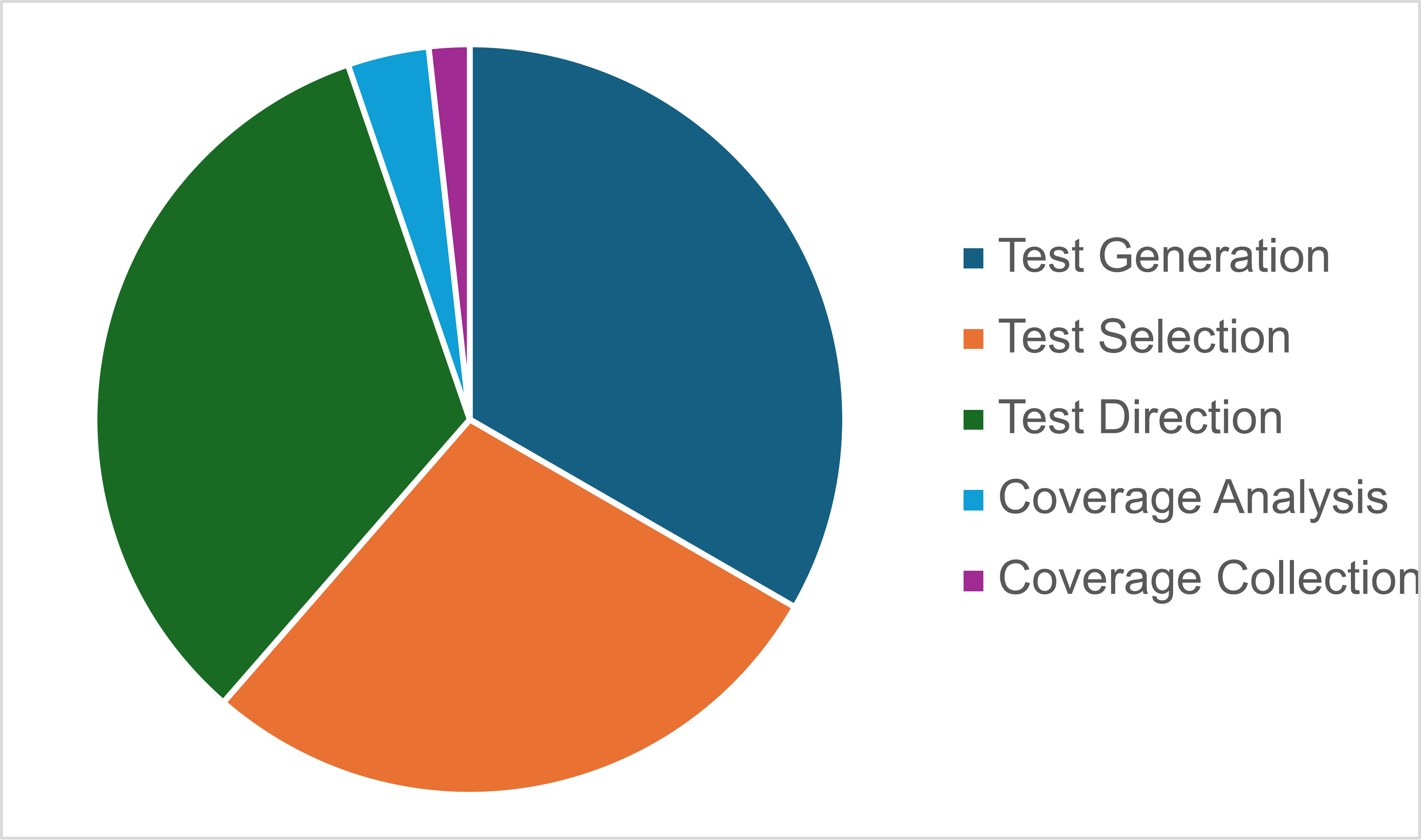}
        \caption{Proportion of techniques by coverage closure activity.}
        \label{fig:verif_covarageclosure_pie}
    \end{subfigure}
    \caption{Left: Proportion of papers by verification activity. Right: Proportion of papers by coverage closure technique.}
\end{figure}

Of the four activities, the majority of papers apply machine learning to coverage closure (Figure~~\ref{fig:verif_process_pie}).  Achieving closure is a significant bottleneck in the electronic design process~\cite{AboelMaged2021}, and the problem of coverage closure can also be framed as a mapping from input to output space for a black box function.  A framing compatible with a wide range of machine learning techniques.  Therefore, it is unsurprising that coverage closure has occupied a significant proportion of the research interest.  

In the research material, the use of machine learning in coverage closure was predominately an even split between test direction where the ML model parameterises a (usually) constrained-random test generator, test selection where the machine-learning selects stimuli from a pre-generated set and test generation where the machine learning generates the stimuli directly (Figure~\ref{fig:verif_covarageclosure_pie}).  
The amount of material for Test Generation relative to Direction and Selection is surprising.  Constrained-random test generators are widely used in industry which facilitates the incorporation of Test Direction based techniques into existing verification environments and workflows. Test Selection is also commonly used to create test sets for regression (periodic testing) and is often widely compatible with different workflows.  However, Test Generation requires domain knowledge to generate legal inputs which is potentially more challenging than Direction and Selection, and it is also potentially more difficult to integrate into an existing verification environment.

Only a small number of coverage analysis and collection related work were found in the sampled literature.  The low representation of these topics may be due to unintentional bias in the sampling methodology.  However,  both activities are associated with large amounts of data in big design projects, something present in industry but more challenging to replicate in an academic research context.  This may explain the lack of academic research material in these areas.

\section{Use Cases, Benefits and Desirable Qualities} \label{sec:usecases}
Using machine learning in verification is applied research with real-world benefits to the electronic design industry.  Progress relies on understanding where machine learning can be applied, what the measures of success are, and how it benefits the verification process.  Research and industry have expressed these as high-level summaries.  However, we found the research to be more granular.  Authors used ML to address specific use cases and measured success against application-specific criteria.  This section uses the sampled literature to collate these use cases and criteria as a platform for future work.  The aim is to provide a qualitative summary of where machine learning is used, what benefits the research aimed to bring, and what the research community views as success in the context of machine learning for dynamic-based functional verification of electronic designs.  We address quantitative (metrics) of success in Section~\ref{sec:evaluation}.

\subsection{Applications for ML in Simulation-Based Testing}
In this context, an application describes a scenario in which machine learning can be used during the verification of microelectronic devices.  It focuses on \emph{what} the practitioner aims to achieve rather than \emph{how} the machine learning can be applied.  The taxonomy in Section~\ref{sec:quantitiveanalysis} (Figure~\ref{fig:mlusetaxonomy}) is based on industry processes and a starting point for practitioners to identify relevant ML research to improve a particular aspect of verification.  While this taxonomy is a quick way to access the literature, there is a range of applications for machine learning within a group such as test generation or selection.

Here, we examine the applications found in the sampled research, emphasising details likely to affect the machine learning solution, including whether inputs are sequential, how machine learning is integrated into a verification process, and what ML is used to predict.  The applications in this section have been synthesised from the sampled literature, and similar applications have been combined only where the loss of detail is unlikely to affect the application of ML.  Conversely, applications have been kept distinct where specific details are likely to affect the machine learning solution.

\begin{table}[t]
    \centering
    \begin{tabular}{|p{7cm}|p{1.5cm}|p{4cm}|}
        \hline
        \textbf{Application} & \textbf{\#Papers} & \textbf{References} \\
        \hline
         Generate inputs to maximise coverage & 20 & \cite{Pfeifer2020, Fajcik2017, Halim2022, Tweehuysen2023, Simkova2015, Dimitrakopoulos2023, Krishna2023, Xia2024, Danciu2022,  Bose2001,  Smith1997,  Gadde2024, Mondol2024, Ohana2023, Elver2016, Wang2022, Shen2008, Samarah2006, Corno2004b, Ravotto2008} \\
         \hline
         Predict input to hit an output & 7 & \cite{Baras2011, Fine2003a, Hsueh2006, Braun2003, AboelMaged2021, Cristescu2021a, Ambalakkat2019} \\
         \hline
         Predict output from an input & 9 & \cite{Ismail2021a, Gogri2022, Ghany2021, Masamba2022b, Guo2010, Masamba2022a, Chen2012, Wang2018, Gad2021}\\
         \hline
         Measure similarity/novelty & 5 & \cite{Liang2023, Zheng2023, Chang2010, Zheng2024, Guzey2008}\\
         \hline
         Improve the quality of coverage & 2 & \cite{Romero2009, Gal2020} \\
         \hline
         Frequently hit the same coverage point / event & 1 & \cite{Moonki2022} \\
         \hline
         Improve the effectiveness of existing methods & 4 & \cite{Stefan2021, Braun2004, Katz2011, Huang2022, Gadde2024}\\
         \hline
         Improve the efficiency of existing closure methods & 6 & \cite{Mandouh2018, Gal2020b, Fine2006, Roy2018, Kumar2023, Phogtat2024}\\
         \hline
         Improve the efficiency of regression testing & 5 & \cite{Guo2011, Parthasarathy2022a, Zachariaova2016, Ikram2017, Jang2022}\\
         \hline
         Generate tests to be reused at a different levels of abstractions & 2 & \cite{Yu2002, Habibi2006} \\
         \hline
         Expose a known bug & 4 & \cite{Thamarai2010, Bhargav2021, Bernardeschi2013, Bernardi2008} \\
         \hline
         Find new bugs & 4 & \cite{Shen2005, Guo2014, Sokorac2017, Wagner2007}\\    \hline
    \end{tabular}
    \caption{The applications of machine learning in simulation-based verification of microelectronic devices.}
    \label{tab:usecases}
\end{table}
Generating inputs to maximise coverage often used reinforcement learning or evolutionary algorithms to create constraints and instruction sequences aimed at increasing coverage. Alternatively, some research uses machine learning to predict test inputs rather than generating them directly. Predicting an input to hit an output is associated with targeting known coverage holes, while predicting an output from an input approaches the problem in reverse, predicting the coverage point hit given a known input.

Machine learning was also used to measure the similarity or novelty between sets of tests. This was common in techniques that identified transaction sequences to simulate from a pre-generated set without coverage information.

Some applications aimed to improve the quality of coverage rather than just the total percentage of coverage points hit. For example, improving coverage evenness by selecting instruction sequences to target infrequently-hit coverage points, and other techniques enhance coverage quality by selecting tests to ensure coverage points are hit from different prior states of the Device Under Verification (DUV).

Although applications of machine learning often result in fewer simulation cycles, some are distinguished by not being standalone methods but instead improving the efficiency of existing methods. Examples include using machine learning to group highly correlated coverage holes and predicting whether an initial state of a device will increase the probability of generating a successful test.  

Applications that improve the efficiency of regression testing are run outside the testing loop and usually have access to information such as design changes and which tests previously detected errors. These applications reduce the number of tests that need to be simulated and some optimise against resource budgets.

Some of the applications relate to improving the effectiveness of machine learning.  For instance, by proposing a communication infrastructure between a DUV and 
 an RL agent~\cite{Stefan2021}, automatically fine-tuning the parameters of a Bayesian Network model leading to better constraints for a test generator~\cite{Braun2004}, or by automatically learning and embedding domain knowledge into a test generator~\cite{Katz2011}.

Bug detection can be split into two types of applications. In the first type, the bug is known, and machine learning is used to find a test sequence that causes the bug to be detected. In the second type, the bug is unknown, and machine learning is used to increase the probability of testing finding bugs.

Research that used machine learning to generate tests to be reused at a different level of abstraction is similar to generative or predictive applications that increase coverage.  However, the aim is not to achieve high coverage per se but to create a test set for use later in development.  For instance, using behavioural simulations written in high level languages to create tests for RT-level or gate-level representations.

The applications in this section are high-level groupings. In practice, a practitioner needs to consider important details specific to their application, particularly in the input and output spaces of their machine learning application.   
In the input space, details to consider include whether the inputs are sequences or singular, how closely related the inputs are to the DUV behaviour (e.g., parameters for a test generator are less closely related than instructions to the DUV), whether the inputs are from simulated or unsimulated tests, and how the inputs are generated (e.g., randomly, expert-written, or from historical information).
In the output space, details include whether the ML model produces a test input (such as a constraint or instruction) or makes predictions about DUV behaviour.

\subsection{Benefits of Using Machine Learning in Microelectronic Design Verification} \label{sec:benefitsofml}
It is common practice for applied research to describe the benefits of a proposed technique.  In this section, we summarise the benefits cited by research against the different machine learning applications.  

We attempted to capture the views of the original authors as closely as possible.  Since benefits are described differently and with a particular focus, it creates overlap.  For example, where one piece of research cites a reduction in the number of simulations, another may cite hitting coverage holes faster or reducing verification time; all of which are related.  We chose to keep this overlap to give a more accurate depiction of the literature.  If research listed more than one benefit, then we listed each separately for the same reason.  See~Table~\ref{tab:benefits}.

\begin{table}[h!]
    \centering
    \begin{tabular}{|p{8.5cm}|p{5.5cm}|}
    \hline
    \textbf{Description} & \textbf{Examples} \\
    \hline
    Reducing the number of simulations and redundant tests & \cite{Cristescu2021a, AboelMaged2021, Ismail2021a, Gad2021, Gogri2022, Stefan2021, Halim2022, Kumar2023, Tweehuysen2023, Dinu2021, Fine2006, Romero2009, Shen2005, Zheng2024, Guo2010, Masamba2022a, Simkova2015, Kamath2012, Liang2023} \\ 
    \hline
    Decreasing simulation time & \cite{Das2024} \\ 
    \hline
    Reducing computational overhead for machine learning & \cite{AboelMaged2021, Gogri2022, Guo2010, Hu2016, Zheng2023} \\ 
    \hline
    Reducing time to reach coverage closure & \cite{Ghany2021, Fine2003a, Fine2006, Guo2010, Simkova2015} \\ 
    \hline
    Reducing verification time & \cite{Gad2021, Ashraf2021, Dinu2021, Fajcik2017} \\ 
    \hline
    Hitting coverage holes faster & \cite{Mandouh2018, Fine2003a, Zheng2023} \\
    \hline
    Reducing expert resources & \cite{Shen2005, Hsueh2006, Masamba2022a, Simkova2015} \\ 
    \hline
    Generalising to different verification environments & \cite{Stefan2021, Mondol2024, Fajcik2017, Guo2010, Simkova2015, Pfeifer2020, Zheng2023} \\ 
    \hline
    Improving ML performance & \cite{AboelMaged2021, Pfeifer2020} \\  
    \hline
    Using verification resources effectively & \cite{Kumar2023, Pfeifer2020} \\
    \hline
    Adding features & \cite{Fine2006, Romero2009, Kamath2012} \\ 
    \hline
\end{tabular}
\caption{Benefits cited by machine learning applications for microelectronic device verification in dynamic-based workflows.}
\label{tab:benefits}
\end{table}

In the context of coverage closure, redundant tests are simulated but do not add to coverage.  More generally, a DUV is simulated for other reasons including generating training data and understanding behaviour.  Since simulating a DUV has a cost in computational and time resources, a large proportion of the machine learning applications cite their benefit as reducing the number of times a DUV is simulated.  This group also includes applications that aim to find the smallest number of transactions to reach an output state~\cite{Stefan2021}.  Applications that decrease simulation time aim to reduce the resource expense of a single simulation rather than the total number~\cite{Das2024}.

Machine learning methods introduce compute cost.  To mitigate this cost, applications cite benefits including reducing training time, the need to retrain regularly, reuse of existing simulation data~\cite{Hu2016}, a low training cost relative to simulation time~\cite{Zheng2023}, and scalable re-training as new training data is generated.

Most research on applying machine learning to coverage closure highlights the benefit of reducing the time to achieve coverage closure. This can be accomplished not only by decreasing the number of simulations but also by shortening the time needed to generate inputs and training data. Reducing verification time was created to encompass applications that report faster coverage closure without specifically mentioning fewer simulations.  

Hitting coverage holes faster relates to techniques that propose to be good at covering hard-to-hit coverage points including methods that create a direct mapping from a coverage point to the input required to reach it.

Reducing expert resources includes applications that reduce the need for human written directives, domain knowledge to set up the technique, and human intervention during coverage closure.

This review finds the research lacks an emphasis on generality.  However, a selection of research cites compatibility with standard UVM environments and different test generators as a benefit.  Approaches that treat the DUV as a black box also cite generality to different DUV designs.

Improving machine learning performance was rarely cited as a benefit, suggesting an emphasis from research on proposing new applications rather than improving existing methods.

A small selection of the sampled material cites the benefits of a proposed technique to operate with constrained resources, such as maximising coverage subject to a time constraint or testing with constrained computing and licenses.

Finally, research also cites the benefits of adding features not necessarily present in a verification workflow. For example, increasing the diversity of inputs to a DUV is one such feature. Another is decreasing the number of cases where a pseudo-random test generator fails to generate a sequence of outputs respecting its constraints. Additionally, increasing the frequency of a single event of interest in the DUV is also cited as a benefit.

The overarching benefit of using ML for verification in the sampled literature is reducing the time spent on verification.  This is motivated by the frequently cited figure of 70\% of design time spent on verification.  However, the time saved by an application may not be realisable in all scenarios.  A device that is quick to simulate relative to the time to generate inputs would not necessarily see the time savings from methods that generate many inputs and simulate only a few.  To encourage generality and the adoption of techniques, we would encourage future research to be specific about the benefits associated with proposed applications.   An approach taken by some authors to aid those adopting their work is to split time into training, simulation and generation.  For practitioners assessing different techniques, we recommend assessing the benefits of each ML approach in the context of their design and verification environment.

\subsection{Qualities of a Test Bench} \label{sec:testbenchqualities}
A test bench is central to a dynamic verification workflow.  The motivation for using machine learning was often seen to enhance an element of a test bench, moving the state of the art closer to the ``ideal''.   Here, we summarise the qualities of a test bench research aims to improve.  

\begin{longtable}{|m{0.15\textwidth}|m{0.7\textwidth}|}
  \hline
    \textbf{Grouping} & \textbf{Criteria} \\
    \hline
    \endfirsthead
    \hline
    \textbf{Grouping} & \textbf{Criteria} \\
    \hline
    \endhead
    \hline
    \endfoot
    \endlastfoot
        Quality & The output is deterministic and repeatable~(\cite{Bergeron2003, Yuan2006} as cited in ~\cite{Wagner2007}). \\
        \cline{2-2}
                & Only valid input sequences to the DUV are generated~(\cite{Bergeron2003, Yuan2006} as cited in ~\cite{Wagner2007}),~\cite{Fine2006}. \\
        \cline{2-2}
                & Transactions stress the interfaces between modules where potential bugs are most likely to be found~\cite{Wagner2007}.  \\ 
        \cline{2-2}        
                & Controls are provided for how often each task is covered using different test directives. \\
        \cline{2-2}
                &  Generated tests are based on the results of previous tests and the requirements of future testing.  \\
        \cline{2-2}
                &  The tester is capable of exhaustively covering the necessary testing scenarios measured via a coverage metric~\cite{Qiu2024}.   \\
        \cline{2-2}
                & The tester can correctly assess whether an output is correct for a given test input~\cite{Qiu2024}.  \\ 
        \hline
        Efficiency & Interfaces seamlessly with existing simulation environment~\cite{Wagner2007}  \\
        \cline{2-2}
            & Tests are ordered to prioritise coverage efficiency at the start of testing and achieving full coverage later in testing~\cite{Baras2011}.   \\
        \cline{2-2}
            & Tests are selected and ordered to cover the task space efficiently~\cite{Fine2006}.  \\ 
        \cline{2-2}        
            & From the first test, each contributes to the verification effort.   \\ 
        \cline{2-2}
            & The tester automatically finds which parameters (from the many in the verification environment) are needed to affect the output to hit a coverage point.  \\
        \cline{2-2}
            & The number of resets required for the DUV over the course of testing is minimised~~\cite{Laeufer2018}.  \\ 
        \hline
        Usability & Engineers have a clear and effective way of biasing a test towards a specific coverage area~(\cite{Bergeron2003, Yuan2006} as cited in ~\cite{Wagner2007}).  \\
        \cline{2-2}
            & Sets of similar inputs (e.g., instructions) are grouped with a short hand notation~(\cite{Bergeron2003, Yuan2006} as cited in ~\cite{Wagner2007})  \\
        \cline{2-2}
            & Tests can be understood in a human readable, simple, test specification language~\cite{Eder2007, Wagner2005}, ~(\cite{Bergeron2003, Yuan2006} as cited in ~\cite{Wagner2007})  \\
        \cline{2-2}        
            & A user is able to configure the tests for either speed or coverage~~\cite{Bernardeschi2013}. \\  
        \hline             
        Functionality & Capability to optimise existing sets of test programs~\cite{Corno2004b}  \\
        \cline{2-2}
            & Generated tests are applicable at both the design stage and post-manufacture to find design faults (bugs) and manufacturing defects. \\
        \cline{2-2}
            & Pipelined processors can be tested where the behaviour is determined by the sequence of instructions and the interaction between their operands~\cite{Corno2004a}.  \\
        \cline{2-2}        
            & The tester infers the relationship between the verification environment's initial state and the generation success of all subsequent instructions in the test~\cite{Fine2006}. \\
        \cline{2-2}
            & Undefined (but necessary) coverage points are identified automatically~\cite{Farkash2014}.   \\
        \hline
        Generalisable &  Minimal human effort and expertise is required to set up and use the test environment  \\
        \cline{2-2}
            & Flexible to verify different design elements*~\cite{Smith1997}.   \\
        \cline{2-2}
            & Flexible to verify different coverage models*~\cite{Smith1997}.   \\
        \cline{2-2}        
            & Flexible to verify at different levels of abstraction*~\cite{Yu2002}.  \\
        \cline{2-2}
            & Easy to verify multiple objectives or at worse to verify for different objectives*~\cite{Bose2001}.   \\
        \cline{2-2}
            & Does not require design specific information beyond that which is available in the design specification~\cite{Wagner2005, Wagner2007, Corno2004b}.  \\  
        \cline{2-2}
            & Test vectors generated at high abstraction levels can be reused to test at lower levels of abstraction to reduce the cost and the overall time for verification and testing~\cite{Yu2002}.  \\
        \hline             
    \caption{The qualities of an ideal test bench for test-based verification and related research papers.  *~denotes without significant rebuilding of the verification environment.}
    \label{tab:qualitiesoftestbench}
\end{longtable}

\newpage
\section{Training and Learning Methods}
\label{sec:training}
Except unsupervised techniques, all methods in the sampled literature required a process of learning to improve the method's performance.  The type of learning fell into one of three categories:
\begin{itemize}
    \item[-] \textbf{Online}: the model learns while it is being used, in some instances, influencing the collection of new data.  
    \item[-] \textbf{Offline}: all training data is available during model creation.  The model is not retrained regularly.  
    \item[-] \textbf{Hybrid}: a small set of training data is used to initialise the model, and new information is regularly integrated during the model's use.
\end{itemize}

\begin{figure}[h]
    \centering
    \includegraphics[width=0.5\linewidth]{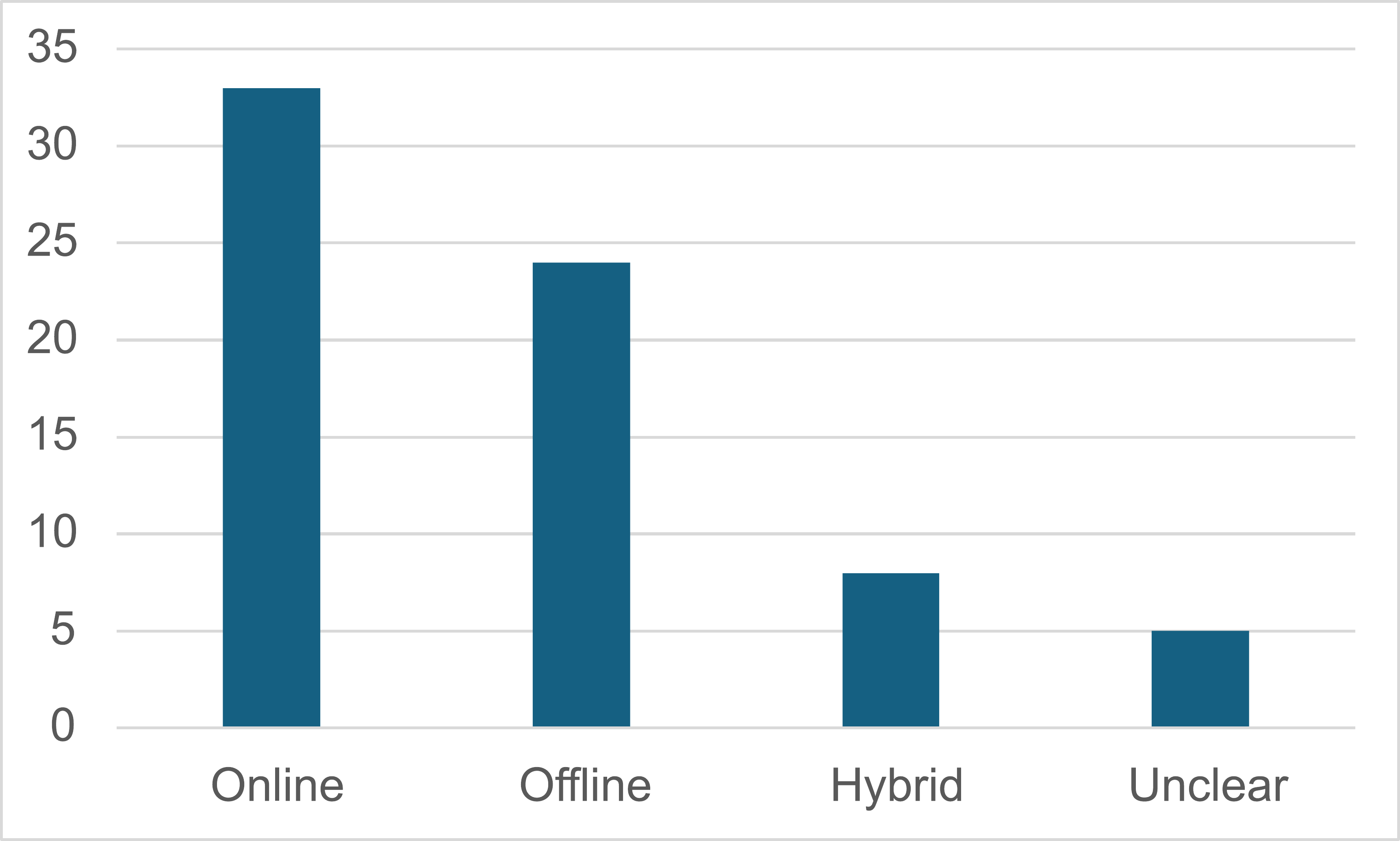}
    \caption{The number of papers by learning type.}
    \label{fig:learningtype_bar}
\end{figure}

Figure~\ref{fig:learningtype_bar} shows the distribution of work by learning type. Online learning is synonymous with reinforcement learning and genetic algorithms that require feedback to guide their learning.  These approaches trade weaker initial performance for the continuous integration of new information.  Conversely, offline learning favours techniques where large amounts of information is available, the cost of errors is high, or training times are long relative to the time to collect new information.  Hybrid learning is a trade-off between online and offline learning.   One example compared online and offline learning, finding online learning had lower overall accuracy, but a lower retraining time made it more scalable compared to offline learning.  

In the literature, many offline learning methods used training data obtained through random based test generation~\cite{Ghany2021}.  Since random-based methods are common in microelectronic device verification, there is likely to be an abundance of this type of data. However, as with other fields of ML, learning requires a balanced, unbiased, dataset.  Randomly generated data sets for a DUV may not achieve this if, for example, some coverage points are hit substantially more regularly than others.  Balancing datasets is discussed, but in general the sampled literature does not examine how information collection may affect the machine learning performance.

Online or hybrid methods retrained regularly in small batches were commonly used when selecting constraints or DUV inputs based on novelty.  Novelty is measured against past examples~\cite{Zheng2023}.  A novel example may not be novel over time after more examples have been seen, necessitating regular retraining to keep the machine learning assessment relevant.  Termed ``concept drift'' in~\cite{Guo2010}, the choice of when to deploy a model and how to retrain can be important.  Once deployed, the learner influences the future examples it will be retrained on, potentially preventing sufficient exploration of the DUV's states to be verified, leading to performance that decreases over time.

Overall, online is the most common learning approach.  In an industrial design and verification process, design changes and continuous production of simulation data mean that all machine learning applications would benefit from integrating new information.  The question is how and when to retrain and any associated trade-off between accuracy and training time.  This question is not commonly addressed in the literature.  Research often frames verification of microelectronic devices as a ``one-time'' learning problem.  A challenge for future research is to move towards solutions suitable for the iterative and rapidly changing designs seen in an industrial setting.

\section{The Use of Machine Learning for Coverage Closure} \label{sec:qualcoverageclosure}
In this section, we discuss coverage models and the application of machine learning techniques to coverage closure.  Coverage closure is the activity of testing all points within a coverage model, and it was the most widely researched verification topic in the sampled literature.

\subsection{Coverage Models}\label{sec:coveragemodels}
Coverage models are derived from a DUV's verification plan.  Points in models represent functionality of interest to the verification team.  A typical project may contain hundreds of these models, and they are typically used to track verification progress.  Coverage closure is reached when the number of verified points (the functionality has been shown to be correct against the specification) passes a threshold.  Achieving coverage closure is one of the conditions for a design going to production.  Research frequently bases an objective function or classification on coverage models.  For instance, a common formulation attempts to learn the relationship between the constraints applied to a random test generator and the coverage points hit.

\begin{figure}[h]
    \centering
    \includegraphics[width=0.8\linewidth]{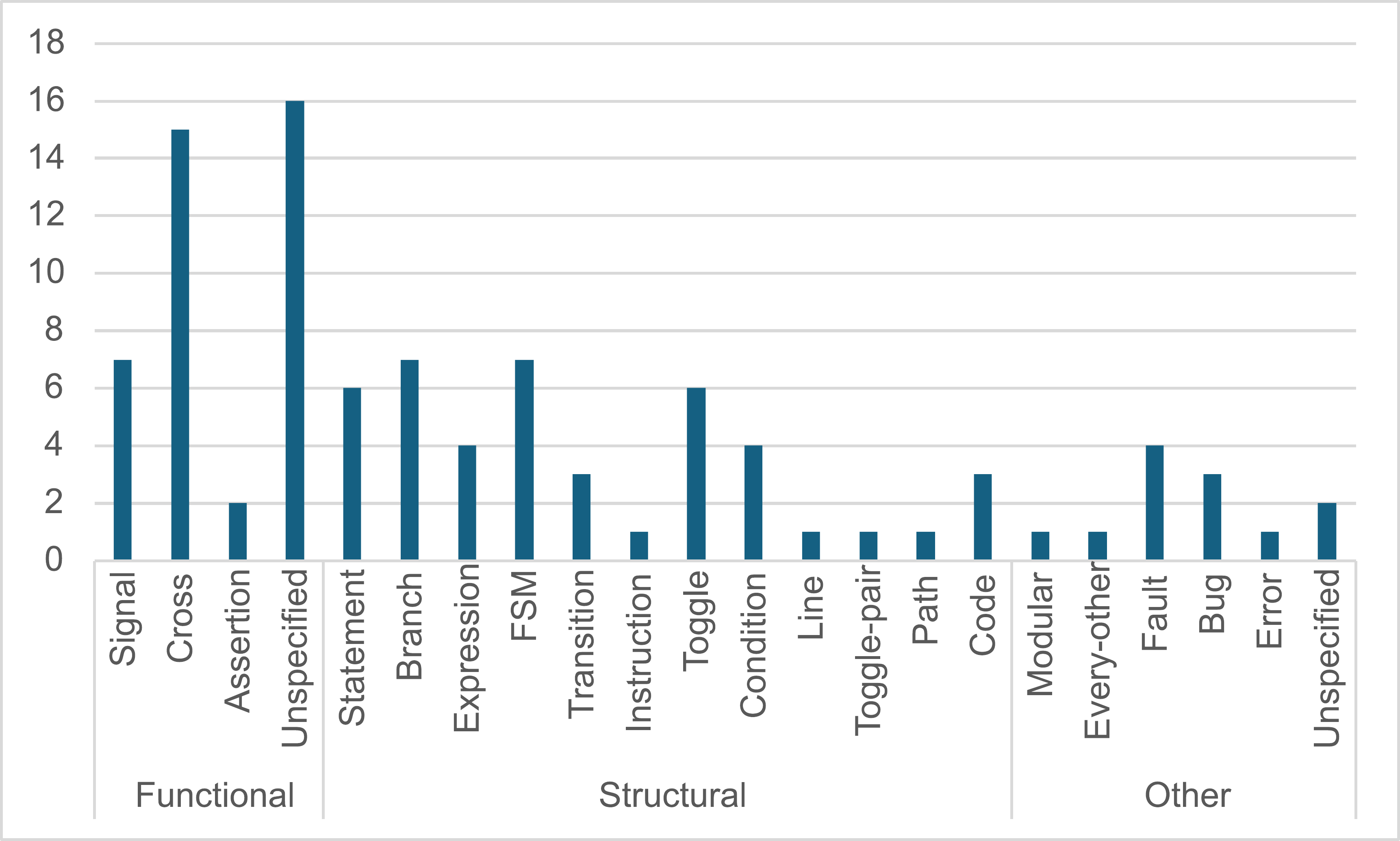}
    \caption{The number of examples found by coverage model.  Where more than one coverage model is used in a single paper, these are listed separately.}
    \label{fig:coverage_bar}
\end{figure}

Given the importance of coverage models in microelectronic device verification, it is unsurprising that approximately 90\% of the sampled literature used a coverage model.  There were two classes of model used (Figure~\ref{fig:coverage_bar}).  Structural models derive automatically from the design and include code (statement, branch, expression), FSM and instruction.  Functional models are created from a DUV's specification and include cross-product and assertion models.  Functional models are commonly created by experts, although there is research into using machine learning (especially large language models) to assist in their creation.  A proportion of work using functional models targeted the range of values for a signal.  For instance, the output of an ALU.  These applications were categorised as ``Signal Values''. 

To preserve information, specialist types of ``models'' not traditionally associated with coverage have been included, where the models are used for a similar purpose.  Bug coverage models are used by works that seek to replicate or test previously identified bugs. Modular coverage ``models'' seek to record the number of cycles a particular module within the DUV is active during simulation.  Their use is seen in papers testing communication devices at the SoC level.  

Three papers used more than one type of coverage model.  Presenting results obtained with multiple types of coverage models helps to demonstrate a technique generalises.  

Several weaknesses were also present in the literature.  Only two examples were seen in the sampled literature of ML applied in conjunction with assertion-based models~\cite{Wang2018, Habibi2006}.  Assertion models are used in both dynamic and static (formal) methods and it is surprising to not find them better represented.  

Functional models were sometimes vaguely described, with 16 out of 40 models in this category described only as ``Functional'' without further qualification of the model.  A clear definition of functional models is important to assess the complexity of the learning problem.  Some authors comment on the relatedness of a coverage model to a DUV's input space, but most do not.  Clear definitions of coverage models are also necessary to enable others to repeat a piece of work.

The number of points in a coverage model (size) may also affect the choice of machine learning, the complexity of the problem and the amount of training required.  A large coverage model often results in a large output space for machine learning. 
However, research did not always give the size of the model.  Approximately one third of the coverage models seen were of unspecified size.  Instead, authors would more commonly describe the coverage as a percentage of the total number of coverage points hit at least once.  Where the size of a model was given, the smallest model had one coverage point representing a FIFO buffer full condition~\cite{Wang2022}, and the largest had 430000 coverage points for an unspecified industrial design~\cite{Mandouh2018}.  The median size the coverage models was 433 for functional models, slightly larger than the 100 for structural models (Table~\ref{tab:coveragemodelsizes}).

\begin{table}
    \centering
    \begin{tabular}{|c|c|c|c|}
        \hline
               & Functional & Structural & Other \\
       \hline
        Median  & 443 & 100 & 33 \\
        \hline
        Maximum & 430000 & 2590 & 10394 \\
        \hline 
        Minimum & 1 & 4 & 4 \\
        \hline
    \end{tabular}
    \caption{The number of points used in coverage models.  Research that either did not use coverage models or did not specify their size is not shown.  Where a single piece of research used different types of coverage model, the size of each is included as a separate value.  Some research uses different models of the same type, for example when applying a technique to different designs.  Where this occurs, the largest and smallest model size is included.}
    \label{tab:coveragemodelsizes}
\end{table}

The size of a coverage model does not necessarily reflect the complexity of using it to train a machine-learning model.  In~\cite{Liang2023}, two DUVs are used with different coverage models, and the authors state one model has coverage points that are harder to hit.  Similarly, in~\cite{Gogri2022}, multiple models are used to optimise coverage closure at the test level.  Two coverage models are subsequently carried forward to optimise at the transaction level because these models were harder to hit.  This discussion about the complexity of the learning problem was rarely seen in the literature but is valuable to anyone applying the technique to a new application.

Demonstrating the generality of a technique requires applying it to different coverage models.  It is unlikely a practitioner would use exactly the same DUV or coverage models as the research.  There are many examples of research that compare different machine learning approaches~\cite{Gogri2020, Gad2021, Braun2003, Gadde2024}, but very few compare a method's performance against different coverage models.

Overall, coverage models were commonly used in the sampled literature.  While some examples exist of research specifying the type of model, its size, and the complexity of relating a DUV's input space to a coverage model, this information is often incomplete or not provided.

\subsection{The ML-Enhanced Verification Environment}
\begin{figure}[h]
    \centering
    \includegraphics[width=0.95\linewidth]{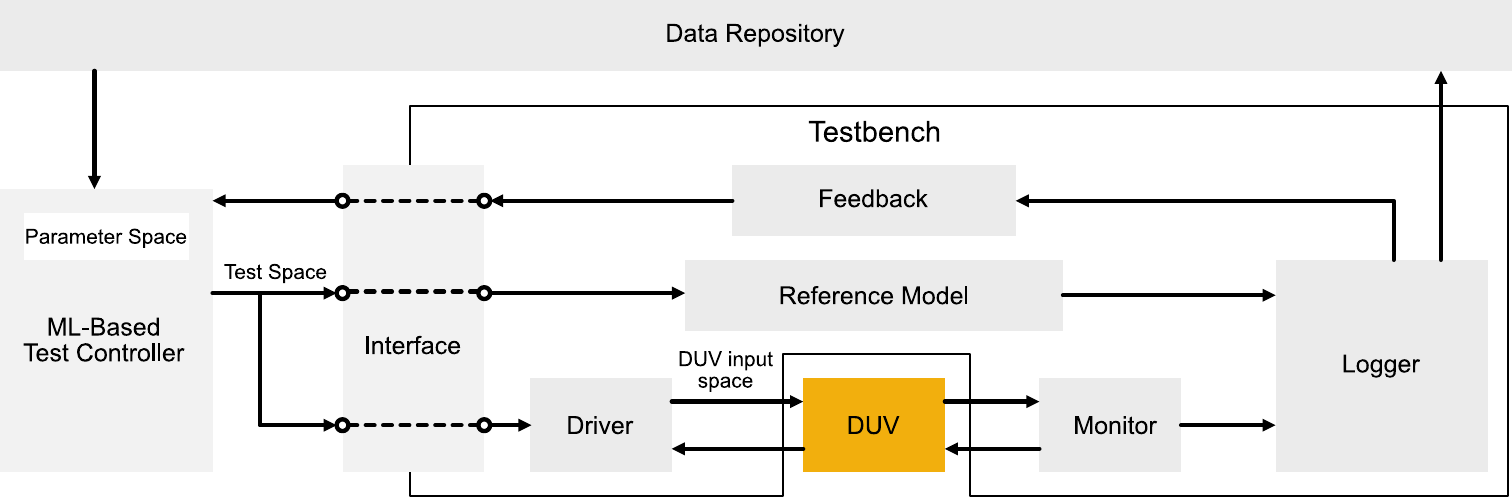}
    \caption{A simplified simulation-based test flow for functional verification using machine learning. Typically, the ML controller supplies tests to the testbench, which can include machine-readable instructions, parameters for a pseudo-random test generator, or bit-level stimuli. It is common for ML-applications to be written in a different environment and require an interface to connect with the testbench.}
    \label{fig:mltestbench}
\end{figure}

Figure~\ref{fig:mltestbench} shows a simplified view of a simulation-based test flow used in ML research for coverage closure.  It modifies the traditional approach (Figure~\ref{fig:generictestbench}) by replacing a human expert with an ML-based test controller.  Generated tests are sent to a simulator and golden reference model.  The simulation drives the DUV to different states and produces outputs that are compared with the reference from the golden model.  During the test, the DUV's states are monitored to record coverage.   Research can be differentiated based on the construction and operation of the ML-based test controller.

Using a random test generator is viewed in the literature as the most basic form of testing, and it is often the baseline against which authors measure the success of proposed improvements.  Instructions are generated randomly, usually with the constraint that only legal instruction sequences are generated.  Given sufficient time, this will in principle cover all states of the DUV and therefore the coverage model, but no guarantees are made on wall-time taken or the distribution of the coverage points hit.  
If random generation is at one end of a spectrum, then in principle, there exists an optimal method at the other end that can find the minimum number of instructions necessary to cover the coverage model with an even distribution across the coverage points.  All the literature in this section proposes a form of test controller that falls somewhere on this spectrum.  Each aims to beat random and come as close as possible to the optimal method.

\subsection{The Application of ML to Coverage Closure}
The applications of ML to coverage closure seen in the literature can be classified based on how the ML-based test controller supplies tests to a testbench (Figure~\ref{fig:mltestbench}).  In test generation, a ML model is used to generate input sequences to a DUV.  For test direction, a ML model is used to enhance the choice of parameters used in an existing generation method (usually a constrained random test generation).  And in test selection, machine learning is used to choose input sequences from a pre-generated set.

ML has been applied to three different input spaces: parameter, test, and DUV inputs.  The parameter space contains the constraints, weights and hyper-parameterises that change the operation of the generation method.  The test space comprises sequences of inputs, and these can be written at different levels of abstraction, including as opcodes or bit patterns. Finally, the DUV input space contains the inputs driven into the DUV and is (usually) represented at the bit level.  Although, there are examples of some behavioural models driving the DUV model with signals at a higher level of abstraction~\cite{Halim2022, Danciu2022}. 

In the following sections, the use of ML is discussed by the its type, where it is applied in the conventional test flow, input space, and abstraction level.

\subsection{Test Generation}
In test generation, a machine learning model creates the inputs that drive a DUV to different states without using an intermediate mechanism such as a constrained random generator.  We found evolutionary and reinforcement learning techniques used to build these test generators.

\begin{table}[t]
    \centering
    \begin{tabular}{|m{10em}|m{8em}|m{12em}|}
        \hline
        \textbf{Type} & \textbf{Sub-Type} & \textbf{References}\\
        \hline
        Reinforcement Learning & - & \cite{Stefan2021}, \cite{Halim2022}, \cite{Ohana2023}, \cite{Gadde2024}, \cite{Mondol2024} \\
        \hline
         \multirow{2}{*}{Evolutionary Algorithm} & Genetic Algorithm & \cite{Smith1997}, \cite{Danciu2022}, \cite{Krishna2023}, \cite{Xia2024}   \\ 
         \cline{2-3}
        
                              & Genetic Program   & \cite{Yu2002}, \cite{Corno2004b}, \cite{Elver2016} \\ 
        \hline
        \multirow{2}{*}{Supervised} & NN* (deep) & \cite{AboelMaged2021} \\
        \cline{2-3}
                                    & NN* (linear) & \cite{Cristescu2021a} \\
        \hline
        Combination & - & \cite{Wang2022} \\
        \hline
    \end{tabular}
    \caption{Use of machine learning in test generation. *Neural Network.}
    \label{tab:testgeneration}
\end{table}

\subsubsection{Machine Learning Types}
\textbf{Evolutionary Algorithms:} Examples of evolutionary algorithms used for test generation are seen from~\citet{Smith1997}'s early work in 1997 to the present day~\cite{Xia2024, Mondol2024}.

Techniques in this area are primarily differentiated by their use of either a Genetic Algorithm (GA) or Genetic Programming (GP) approach.  The difference between the two is subtle in the case of test generation.  
Both GP and GA generate instructions, but GP evolves a program with structures like loops and branches, while GA evolves an array of instructions.  For instance, GP approaches reviewed used directed graphs to represent the flow of a program~\cite{Elver2016, Corno2004b}, or a sequence of inputs to a DUV~\cite{Yu2002}.  In works using a GA, the encoding used was an array representing a sequence of inputs over time~\cite{Smith1997, Danciu2022, Krishna2023, Xia2024}.  

The inputs forming a genome in GA approaches range from low-level bit representations of opcodes, addresses and immediate values in~\cite{Xia2024}, to high-level representations such as assembly code instructions to verify Cache Access Arbitration Mechanism in ~\cite{Smith1997} or a set of boolean's indicating whether a message is sent between two addresses during Network-on-chip communication~\cite{Krishna2023}.

In addition to the use of GP or GA and the encoding, the choice of algorithm was also a distinguishing feature.  We found limited variety in works using GPs (two out of the three used the $\mu$GP approach described in \citep{Corno2004b}).   
Greater variety in the algorithm was seen amongst works using GAs, specifically in how the selection and mutation operators were defined.  This reflects the need to maintain legal encoding of genomes following an operator, and this requirement varied by applications.  
All work using EAs for test generation used a fitness function based on coverage to guide learning.  However, these works differed in the complexity of this calculation.  Some fitness functions were based on simple measures such as statement coverage~\cite{Yu2002}, whereas others, predominately used for fault detection (Section~\ref{sec:faultdetection}), used multi-objective measures combining structural coverage models of State, Branch, Code, Expression and Toggle~\cite{Ravotto2008}.

Despite work in this area being differentiated by the choice of algorithm, how the test sequence is encoded, and the fitness function used, we found no discussion of the effect of each on the learning and its relative success. For instance, encoding as a graph enables the algorithm to operate on loops and jumps, whereas genome representations are limited to operating on a sequential array.  Encoding as bit-level inputs gives a high level of control but the algorithm operates on a low level of semantic meaning.  These decisions about how the evolutionary algorithms are applied is likely to affect learning, but there's currently insufficient research to conclude their effect on coverage-closure.

\textbf{Reinforcement Learning:} The use of reinforcement learning (RL) to generate input sequences to a DUV has only been studied recently compared to Evolutionary approaches (Figure~\ref{fig:mltypetimeline}).  RL has been demonstrated on small designs for functional coverage including an ALU~\cite{Stefan2021} and LZW Compression Encoder~\cite{Ohana2023}, and later works have applied RL for (structural) code coverage of a RISC-V design~\cite{Gadde2024}.  We found no examples of research which used reinforcement learning for \emph{functional} verification of a complex device at the level of a microprocessor.  We view RL as the least proven of all the techniques surveyed for coverage closure.

RL has, in principle, properties that make it suited to test generation~\cite{Halim2022}.  It acts to maximise total cumulative reward over a sequence of state-action pairs.  Unlike supervised learning, it changes the state of the DUV, receiving immediate feedback which is used to inform its next action, and potentially avoiding sequences which do not add to coverage.  Unlike evolutionary learning, it acts sequentially enabling greater control over the input sequence.  Also, digital designs are inherently compatible with Markov Decision Processes, a representation used by modern RL techniques.  A digital design can be represented as an FSM, where a state is completely described by the DUV's current combinational and memory elements.  Therefore, digital designs satisfy the Markovian property~\cite{Shibu2021}.

One of the challenges for RL is that coverage may be insufficient information to guide learning.  For example, a rare event or coverage hole may generate rewards too spare to guide the learning in a reasonable time~\cite{Shibu2021}.  For example to trigger rare assertions, in \cite{Mondol2024}, one of the actions circumvented the reward signal and chose a test pattern found through static analysis of the code to target RTL code lines.  A solution for white-box testing is to build the reward signal with additional monitors placed on internal signals, similar to that used in~\cite{Wagner2005}.  There are also RL approaches for sparse rewards environments, but these were not seen in the sampled literature.

There is also the challenge of a large actions space.  In \cite{Mondol2024} the solution was a set of actions which mutated the previous test pattern, limiting the action space but potentially encumbering the agent if the current test pattern (and it's variants) place the agent on a poor trajectory.  In ~\cite{Ohana2023}, the DUV was limited to 4-bit inputs to create an action space of 16.

\subsubsection{Benefits of ML for Generative Techniques} 
More generally, we see benefits to using ML for test generation.  The benefit of generative techniques is greater control over the test sequences than directive or selection techniques.  This control may enable results closer to an ideal coverage curve.  We found no examples in the literature which investigated this point.  However, the literature suggests application for ML-enhanced test generation tied to edge cases where the level of control is beneficial.  For instance, in functional coverage for an LZW encoder where input sequences are very specific~\cite{Ohana2023} and random generation hit only 28 out of 136 coverage points, and in~\cite{Shibu2021} where RL was rewarded for finding rare events in an RLE compressor.   In~\cite{Halim2022} RL was only beneficial in complex signalling scenarios where constrained-random struggled to achieve coverage.  In this respect, the use of generative techniques is currently similar to formal techniques.  Greater complexity and resources are balanced by their capability for coverage in edge cases.  Unlike formal methods, RL and EA in principle scale to complex designs, evidenced in~\cite{Mondol2024} where an RL approach was able to find inputs to break security assertions where an industrial grade formal tool failed due to the complexity of the design.     More research is needed to understand the trade-offs.

\subsubsection{Challenges for Using ML to Generate Tests} 
A challenge when using ML with test generation is interfacing the machine learning elements with test benches written in languages that do not natively support ML functions.  In \cite{Xia2024}, a GA is wrapped into a UVM framework to create a standardised architecture usable with different DUVs.   The challenge of interfacing ML techniques with existing test benches for test generation is more acute for RL because most authors used it to generate instructions in the loop with the DUV, thus requiring feedback after each instruction is processed.  Authors using RL techniques interfaced models written in Python, with test benches written in hardware description languages such as SystemVerilog, and each presented architectures to enable a two-way flow of information on a per cycle basis.  In~\cite{Shibu2021}, an open-source library to allow RL-driven verification written in Python to interface with an existing SystemVerilog test bench is presented.

A further obstacle to using ML for test generation outside specific cases is the requirement to generate legal test sequences.  Sequence legality is domain knowledge and there's a question of how the ML acquires it.  In the works using RL, authors defined the problem or the actions the ML could take such that any input sequence it generates was legal.  For instance, applying RL to an ALU~\cite{Stefan2021} or a LZW compression block~\cite{Ohana2023} which accepts any combination of inputs.  We did not find an RL example where learning the domain knowledge for legal sequences was included in the learning.

In EA approaches, the requirements for legal instructions were encoded in the genetic operators.  For instance, in~\cite{Xia2024}, constraints are placed on the location of cross-over operations to prevent invalid instructions from being created.  

Restricting the problem to IP blocks that accept any input, while providing a valuable proof of concept, can be toy problems often not relevant to industry~\cite{Halim2022}.  These block-level toy problems are at a level of complexity where a static analysis tool such as a SAT solver would be able to verify formally with an assurance of fully exploring the coverage space.  A guarantee that stochastic machine learning techniques cannot give. 

There are further challenges to using ML techniques for test generation in the EDA industry beyond demonstrating their capability to learn legal instruction sequences.   Firstly, there is a resource cost to learning domain knowledge which may already be known to the verification engineers.  Secondly, all examples in this review generated instructions to accelerate coverage closure for a specific version of a device.  This means re-training may be required for each device change or when starting a new project.  Thirdly, all the techniques required parameterisation by an expert.  For instance, in~\cite{Halim2022}, hyper-parameters including the episode length, number of episodes, neural network depth and layer width were manually chosen.  Fourthly, the techniques researched for test generation are guided by reward or fitness functions.  Some authors regard these ``objective functions'' as how verification engineers can focus the generation to areas of interest~\cite{Shibu2021}, but most of the material surveyed based these functions on coverage.  Using coverage models reduces the need for additional expertise beyond the existing verification process.  However, some coverage models with hard to hit coverage points may give sparse feedback to the learner, and it's unclear whether generic reward/fitness functions would work in all cases.  Arguably, if a verification engineer is required to create fitness/reward functions to target the model’s output then the use of ML is shifting the design effort from writing test cases to setting ML models.  This is undesirable unless a substantial time saving could be shown.  Finally, the high cost of setting up current ML test generation techniques is especially evident at low coverage percentages.  Both EA and RL techniques use stochasticity to explore the solution space (particularly at the start of training) and have been shown to perform no better than random stimulus~\cite{Halim2022} until coverage increases.  There is an argument to be made that the stochastic exploration of these methods at low coverage may be of higher quality (from a learning perspective) than random generation, resulting in a better solution overall than techniques explored in the next section that use a randomly generated dataset with supervised methods.  However, no research was found investigating this point.

Cumulatively, these reasons lead to a lack of generality, a need for specialist expertise, and high training costs, creating a barrier to industrial adoption.
Applying ML to test direction instead of generation is a popular alternative which lowers the learning cost by removing the need to learn how to generate legal test sequences.

\subsection{Test Direction} \label{sec:testdirection}
We use Test Direction to describe applications that use ML to direct a piece of apparatus to generate test sequences. 

Within Test Direction, we found works either targeted single hard-to-hit coverage holes or attempted to direct coverage to efficiently hit many coverage points. Bayesian Networks were an example of the former, after training they could be interegated to find the constraints most likely to hit a coverage point.  GAs which structure the learning by changing the fitness function are an example of the latter, the learning drives the random-test generator to hit different coverage points.  

Compared to Test Generation, a wide variety of \emph{supervised} machine learning techniques have been applied to Test Direction including Bayesian Networks~\cite{Fine2003a}, Inductive Logic Programming~\cite{Hsueh2006}, and Neural Network based techniques~\cite{Fajcik2017, Gal2020}.

\begin{table}[t]
    \centering
    \begin{tabular}{|m{10em}|m{12em}|m{14em}|}
        \hline
        \textbf{Type} & \textbf{Sub-Type} & \textbf{References}\\
        \hline
        Evolutionary Algorithm & Genetic Algorithm &  \cite{Bose2001}, \cite{Samarah2006}, \cite{Habibi2006}, \cite{Shen2008}, \cite{Simkova2015} \\
        \hline
        \multirow{5}{*}{Supervised} 
                   & NN* (recurrent) & \cite{Fajcik2017} \\
                   \cline{2-3} 
                   &  Bayesian Network  & \cite{Fine2003a}, \cite{Braun2004}, \cite{Fine2006}, \cite{Baras2011} \\
                   \cline{2-3} 
                   & Inductive Logic Program & \cite{Hsueh2006} \\
                   \cline{2-3} 
                   &  Comparison  & \cite{Braun2003}, \cite{Ambalakkat2019} \\
        \hline
        Reinforcement Learning & - & \cite{Gal2020}, \cite{Pfeifer2020}, \cite{Huang2022}, \cite{Tweehuysen2023} \\
        \hline
        Mixed & - & \cite{Kumar2023} \\
        \hline
    \end{tabular}
    \caption{Use of machine learning in test direction.  *Neural Network.}
    \label{tab:testdirectionpapers}
\end{table}

\subsubsection{Machine Learning Types}
\textbf{Bayesian networks (BN)} were a popular technique for test direction in the 2000s (Figure~\ref{fig:mltypetimeline}), with early work in~\cite{Fine2003a} and~\cite{Braun2003}.  A BN is a graphical representation of the joint probability distribution for a set of random variables.  When used for test direction, these variables are parameters for a test generator (inputs), elements of a coverage model (outputs), and hidden nodes for which there is no physical evidence but (by expert knowledge) link inputs to output.   An edge represents a relationship between two random variables.  The network topology represents the domain knowledge of how test generator parameters relate to coverage.  A fully connected network represents no domain knowledge~\cite{Fine2003a}.  Typically, authors divide the creation of a BN into three steps: define the topology, use a training set to learn the parameters of each node's probability distribution, and interrogate the network to find the most probable inputs that would lead to a given coverage point.  The ability to directly predict constraints needed to hit a coverage point gives the approach its power.  However, a frequent criticism was the expertise and time required by a human to create the network topology, thereby limiting scalability and generality.    In~\cite{Fine2006}, these criticisms were addressed using techniques which automatically created the Bayesian network, with later work by~\cite{Baras2011} to further assist their creation.  Although Bayesian reasoning remains popular, the work on artificially created Bayesian networks appears to have stopped after~\cite{Baras2011}, with research interest switching to other techniques, including decision trees and neural networks.  No research was found exploring how the inference power of BN compares to these other approaches, particularly for coverage points where there is no evidence (coverage holes).

\textbf{Genetic algorithms} were also a popular technique for test direction prior to the rise of interest in supervised techniques.  In~\cite{Bose2001}, a GA is used to target buffer utilisations for a PowerPC architecture, in~\cite{Samarah2006} simplified models of a CPU and Router are used, and in~\cite{Simkova2015}, an ALU and Codix-RISC CPU is verified against structural and functional coverage models.  The integration of a GA into UVM architecture is discussed in~\cite{Simkova2015}.

In test direction, a test generator produces many test programs and corresponding coverage hits for a single instance of input parameters (directives).  We see authors structuring the learning by shaping the GA’s fitness function to achieve coverage across multiple objectives.  In~\cite{Bose2001}, the directives to hit two objectives were evolved by first basing fitness on an 80:20 split for the two objectives, then changing to 50:50 once the first objective was met.  In~\cite{Samarah2006}, a four-stage fitness function was used which initially targets all coverage points at least once and then moves to target minimum coverage over four stages.

Authors derive the chromosome encoding directly from the parameter space of the generator, and because each generator has a different input space, there is no single ``right'' encoding to use.   In~\cite{Samarah2006}, the encoding is based on splitting probability distributions for each directive into cells and evolving the weight and width of each cell.
The importance of how generator directives are encoded into a genome was also highlighted in~\cite{Bose2001}, finding that encoding the biases into a structure improved the max buffer utilisation vs random organisation.  This raises a difficulty in using GAs for test direction.  Encoding affects the coverage closure performance, but each test generator has a different parameter space.  Therefore, a practitioner would need to find a good encoding for each test generator used.  Whether or not a universally ``good'' encoding exists for constrained-random test generators remains an open question.

Despite the success of GAs, the large number of parameters and expertise to setup a GA remains a blocker for their use in industry for test direction.  We did not find work which researched the generality of their solutions, suggesting that the evolutionary process would need to be rerun for each coverage model and design change.

\textbf{Supervised Learning} Supervised techniques are trained on labelled data.  The majority of work generates the training set based on the results from random test generation. 
We also see authors proposing approaches to reduce the size of the training set, such as a implicit filtering used in~\cite{Gal2020}.

The abundance of labelled data during dynamic-based verification and the need to lessen the expertise and setup cost seen in other types of ML may explain the recent research interest in supervised techniques for test direction (Figure~\ref{fig:mltypetimeline}).  Different base functions and techniques have been researched including neural networks, Bayesian networks and logic programs (Table~\ref{tab:testdirectionpapers}).  Applications seen range from block level IP, such as a comparator~\cite{Ambalakkat2019}, to complex devices including a powerPC pipeline\cite{Braun2003}, RISC core~\cite{Fajcik2017} and five-stage pipelined superscalar DLX processor~\cite{Hsueh2006}. 

One approach seen is to train a model to predict the mapping between constraints and coverage points~\cite{Ambalakkat2019}.  Another is to predict the number of times to repeat a randomly generated test~\cite{ Kumar2023}, and in~\cite{Fine2006}, relate the initial state of the DUV to the generation success.  The variety of techniques and applications seen in the research suggests the flexibility of supervised techniques and suitability for test direction. However, all the supervised techniques found required parameterisation (as with GAs and Bayesian networks), so despite the recent interest, there remain the issues of generalisation, and the expertise to set up the learning. 

Each test simulated on the DUV creates new labelled data relating the input parameter space to the coverage points hit.  As discussed in~\cite{Gal2020}, supervised methods make trade-offs based on how the generated data is used.  First is the quantity of training data to acquire before using the ML model.  A model trained on a small training set is likely to produce poor prediction at first but improve faster by reducing the probability of covering the same points.  The trade-off is more time spent retaining the model as new data is generated.  The second trade-off is the order coverage points are targeted.  Targeting easier-to-hit coverage points at the start can achieve faster progress during early verification.  Hard-to-hit points are then targeted later when more labelled data is available and the ML model is more mature.  Alternatively, targeting hard-to-hit points during early verification (assuming they're known) may fail but still advance coverage by hitting easier-to-hit points.

\textbf{Reinforcement learning} has had success in learning sequences of actions for complex functions where its actions are high level compared to the process they interact with (cite examples of Alpha Go, Atari Games etc).  It is perhaps surprising that we found few examples of their use in Test Direction.  One reason for this is the complexity of setting up the learner.  Notably, each example for using RL with Test Direction used a different algorithm and framing of the problem. The problem of choosing constraints is framed as a Gaussian process multi-arm bandit problem in~\cite{Huang2022}, and an upper-confidence bound approach is used to balance exploration vs exploitation when selecting which constraints to pick next. In~\cite{Pfeifer2020}, the problem is framed as a hidden Markov model and uses a Raindow RL agent.  Finally, in~\cite{Tweehuysen2023}, the actions are constraints, cover points are states, and an actor-critic approach is used to train the RL agent.  

Reinforcement Learning (RL) has the potential to outperform other methods.  In~\cite{Pfeifer2020}, an RL algorithm achieved slightly higher coverage in less time than an existing Genetic Algorithm (GA) method.  However, this is the only example found in the sampled literature that compares RL to other machine learning methods.   It remains an open question whether the additional cost and complexity of setting up an RL agent are justified by its potentially better performance for test direction.

\subsubsection{Benefits of using ML to Direct Testing}
In test direction, the ML does not generate instructions.  This can circumvent many of the difficulties associated with generating legal instructions.  It also enables domain knowledge to be embedded in the test generator, thereby reducing the size of the learning task.  For instance, knowledge about which sequences of instructions and addresses create edge cases is more likely to uncover errors in a design.  The reliance on a separate generator also makes it easier to interface the machine learning with existing test benches, with communication between the two occurring at the level of constraints that otherwise would have been written by an expert.

\subsubsection{Challenges for using ML to Direct Testing}
Machine learning faces a number of challenges when used to direct a device to generate tests.  Firstly, feedback on the coverage achieved by a set of test directives occurs after the generated test sequence has been simulated on the DUV.  Compared to Test Generation, feedback is slower, and the learner must wait until the end of the complete test sequence to see the results.

Secondly, industrial generators used for constrained random testing may contain thousands of parameters.  A learner must identify those needed to cover a particular model.

Thirdly, a general challenge for using ML for coverage closure is to infer inputs needed to cover holes, creating a particular challenge for supervised techniques.  A hole, by definition, does not appear in a training set.  Unlike GAs and RLs, the supervised techniques seen here are not ``active learners'' in the sense they cannot explore a space, instead relying on the training examples presented to them.  Therefore, supervised techniques place greater reliance on the inference power of the model.  There is limited research which compares different model types.  
In~\cite{Braun2003}, the performance of a Bayesian network is compared to a tree classification technique.   However, no research was found that directly investigated how the choice of model affected inference power for unseen examples. 

Fourthly, a challenge for supervised techniques is creating high-quality training data.  Training sets produced by random sampling are not guaranteed to provide an even spread of examples across the coverage space.  Usually, the reverse is true, and these randomly produced data sets have many examples of easy-to-hit points and very few of the hard-to-hit points.  Some authors attempt to combat this deficiency by shaping the training set. 

Lastly, in the case of constrained random test generators, the output produced for a set of parameters is random.  This stochasticity creates a probabilistic relationship between the input and coverage spaces.  Therefore, the machine learning technique is required to learn from probabilistic relationships.   These relationships are often more challenging to learn and require more training examples.

\subsection{Test Selection}
In constrained random approaches,  some tests do not add to coverage and can be considered redundant.  Test selection is a technique which aims to reduce simulation time by filtering out redundant tests before they are run on the DUV.  The research in this section does this \emph{during} verification testing, which makes it distinct from techniques which run offline and aim to create an optimal test set for regular regression testing. In principle, test selection can reduce verification time when it is cheap to generate but expensive to simulate sequences of instructions on a device.

\subsubsection{Machine Learning Types}
Research in test selection techniques can be split into two types based on whether knowledge of coverage is required.  In the first type, tests are selected based on their similarity to previously simulated tests.  This requires a measure of similarity but does not require knowledge of coverage.  The assumption is that input sequences sufficiently dissimilar will hit different coverage points.  Since coverage data is not required, research has focused on unsupervised learning, using a one-class SVM.

The second type of test selection technique learns a relationship between a test input and coverage.  It uses this information to predict the likelihood a new test input will add to coverage.  For instance, \citet{Guo2010} uses a two-class SVM to select tests for full functional verification of a RISC processor (Godson-2).  The disadvantage of this approach is that it requires simulating some redundant tests to initialise the machine learning model.  However, it makes no assumption about the relationship between input similarity and coverage.   

A test selected without knowledge of coverage will subsequently generate coverage data relating the input and output spaces of the DUV.  This has led researchers to combine both test selection techniques in the same verification workflow.  \citet{Masamba2022a} describes an approach that combines coverage with novelty-directed test selection to contribute to the verification of a commercial radar signal processing unit. 

Interest in novelty detection extends outside of functional verification.  This interest has created different approaches and approximately 40\% of the work reviewed in this section compares two or more techniques.    \citet{Zheng2023} compares the use of an Autoencoder, counting unactivated neurons, and a technique which automatically generates labels to score tests based on coverage.  \citet{Ghany2021} investigates neural-network-based techniques and compares them to using an SVM and decision trees.  

\begin{table}[t]
    \centering
    \begin{tabular}{|m{10em}|m{10em}|m{16em}|}
        \hline
        \textbf{Type} & \textbf{Sub-Type} & \textbf{References}\\      
        \hline
        \multirow{2}{*}{Supervised} &  SVM*  & \cite{Romero2009}, \cite{Guo2010}, \cite{Chang2010}, \cite{Chen2012} \\
        \cline{2-3}
                    & NN (deep) & \cite{Wang2018} \\
        \cline{2-3}
                    &  Comparison  & \cite{Zheng2023}, \cite{Gad2021}, \cite{Masamba2022b}, \cite{Zheng2024}, \cite{Ghany2021}, \cite{Gogri2022}, \cite{Liang2023}, \cite{Ismail2021a} \\
        \hline
        Unsupervised & SVM* & \cite{Guzey2008} \\ 
        \hline
        Mixed & - & \cite{Masamba2022a} \\
        \hline
    \end{tabular}
    \caption{Use of machine learning in test selection.  *Support Vector Machine.}
    \label{tab:testselection}
\end{table}

\subsubsection{Benefits of Using ML to Select Tests}
Compared to test direction and generation techniques, test selection can be the easier to integrate with existing verification environments.  While there is evidence to suggest using coverage data can further reduce the number of simulated tests required to achieve coverage, a test selector which filters tests based only on the similarity of the input space has been shown to be effective; and does not require online learning or changes during a project.  Given the wider interest in novelty detection within machine learning, and the EDA industry's familiarity with test selection for regression optimisation, there is space for more research in test selection

\subsection{Level of Control}
In general, the challenge of learning a relationship between input and output spaces depends on how abstract these spaces are compared to the underlying process that connects them.  Abstraction is a part of the conventional EDA design process.  Electronic hardware design creates models at different levels of abstraction, from behavioural to gate level.  There is also research to reduce the cost of test generation by reusing tests at different levels of abstraction.  For example, to automatically translate a test created at behavioural to gate level~\cite{Yu2002, Habibi2006}.  In this section, we discuss the implications of the level of abstraction to the application of ML to coverage closure.  The aim is to provide a practitioner with a granular means to discriminate between research on this topic.

In the ML-based verification environment (Figure~\ref{fig:mltestbench}), three spaces are identified:  parameter, instruction, and test, and each space can be represented differently (Table~\ref{tab:spaces_to_abstractions}).  Inputs to a DUV are also described at different levels of abstraction, for instance, bit pattern (machine code), opcode and operand (assembly language), constraint, and signal value in a behavioural model (e.g., a traffic light controller), creating a wide range of options.
\begin{table}[h]
    \centering
    \begin{tabular}{|m{12em}|m{12em}|}
    \hline
    \textbf{Space where ML is applied} & \textbf{Representation of the data} \\
    \hline
    Input parameter   & Constraints, random seed or hyper-parameters \\
    \hline
    Instruction & Opcode, signal value, or bit pattern\\
    \hline
    Test & A test identifier, graphical representation of test sequence\\
    \hline
    \end{tabular}
    \caption{Examples of the abstractions used in machine learning for the verification of electronic hardware}.
    \label{tab:spaces_to_abstractions}
\end{table}
Research was found to apply machine learning to control one of these spaces at a specific level of abstraction.  For instance, learning to control the instruction space at either the opcode level or bit level.  Since these spaces and levels of abstraction are relatable to the same low-level design, this creates a choice for how to apply machine learning to achieve coverage closure.

A key question to consider is how the choice of space and level of abstraction affect the complexity of learning and the effectiveness of the machine learning model to speed up verification.  However, we found very little material which sought to answer this question.  \citet{Gogri2020, Gogri2022} investigated the difference between filtering test stimuli at the instruction and constraint levels, finding that the machine learning applied at the constraint level was effective when the input space (constraints) and output space (coverage) were closely related.  However, machine learning applied at the instruction level was more effective when this relationship was more complex.  

From a learning perspective, the state space is smaller at higher levels of abstraction.  A smaller state space may make learning easier, but the relationship between high-level instructions and low-level features may be less direct.  Other authors highlighted that writing tests at high levels of abstraction and translating to the hardware level via a compiler may not be as successful as tests written at the hardware level.  Compiler optimisations and strategies prioritise efficient input sequences.  Therefore, these may not use the full range of all possible instructions and addressing modes~\cite{Smith1997}.

The choice of space and abstraction level is equivalent to feature selection, a crucial part of the success or otherwise of machine learning applications.  Some research in coverage closure has attempted to automate feature selection, but the topic is under represented in the EDA literature.

\subsection{The Use of Machine Learning for Coverage Collection and Analysis}
\begin{table}[h]
    \centering
    \begin{tabular}{|m{10em}|m{8em}|m{12em}|}
        \hline
        \textbf{Type} & \textbf{Sub-Type} & \textbf{References}\\
        \hline
        Coverage Analysis & Supervised & \cite{Mandouh2018},  \cite{Gal2020b} \\
        \hline
        Coverage Collection & Combination & \cite{Roy2018} \\
        \hline
    \end{tabular}
    \caption{Use of machine learning for coverage analysis and coverage collection.}
    \label{tab:coveragecollectionanalysis}
\end{table}

Dynamic-based test methods typically generate large amounts of coverage related information.  Where the majority of techniques seen used coverage data to either directly or indirectly choose stimuli for the DUV, a small number of techniques took a different approach.   

Collecting coverage data adds a computational overhead when simulating a design.  The test-bench must monitor the relevant elements of a design via a scoreboard to record how often a coverage point is hit.  Large coverage models increase this overhead causing simulations to take longer.  In~\cite{Roy2018}, k-means is used to select a small subset of the design to collect coverage, and DNNs predict the coverage of the rest of the design from this small subset.  The author's claim this approach complements existing practice where regressions with full coverage collection are still run, but the technique enables a prediction of coverage in-between those full runs using less computational overhead.  

Two examples were found using machine learning to exploit the relatedness of coverage points to reduce simulation time. Both apply the principle that, when a test hits a coverage point, it has a high probability of also hitting nearby coverage points.  In~\cite{Gal2020b}, clustering techniques using k-means and heuristics are used to identify coverage holes by grouping similar holes together and find a coverage point to target the group.  The approach assumes that related coverage points have similar textual names.  A similar approach is used in~\cite{Mandouh2018}, except similarity between coverage points is based on Jaccard similarity and euclidean distance.

\section{The Use of Machine Learning For Bug Hunting}
\label{sec:bughunting}
\begin{table}[h]
    \centering
    \begin{tabular}{|m{10em}|m{8em}|m{12em}|}
        \hline
        \textbf{Type} & \textbf{Sub-Type} & \textbf{References}\\
        \hline
        Supervised & - & \cite{Moonki2022}, \cite{Shen2005} \\
        \hline
        Evolutionary Algorithm & Genetic Algorithm &  \cite{Bhargav2021}\\
        \hline
        Reinforcement Learning &  &  \cite{Wagner2007} \\
        \hline
        Combination & - &  \cite{Guo2014},  \cite{Sokorac2017}\\
        \hline
    \end{tabular}
    \caption{Use of machine learning for bug hunting.}
    \label{tab:bughuntingpapers}
\end{table}

In the literature, a small number of authors made a distinction between coverage closure that aims to measure verification progress against the DUV specification and bug hunting that attempts to replicate conditions expected to find bugs.  A comparable with existing practice is where an expert writes a test program to target a small number of challenging DUV states.   In~\cite{Wagner2007}, this is described as ``stress-testing'' where a Markov model represents machine instructions and feedback from signal monitors in a design are used to update transition probabilities.  Over time, the instruction sequences to excite signals of interest are generated more often.  In~\cite{Moonki2022}, an approach using linear regression is described to replicate the conditions for a deadlock to occur, and in~\cite{Shen2005}, a neural-network is trained to select constraints for a test generator to hit pre-defined bugs.  The constraints were written by an expert.

The approaches described above assume knowledge of where bugs are most likely to occur in a design.  An alternative approach is described in~\cite{Guo2014}.  Machine learning is used to predict bugs in designs based on historical data from design revisions.  A genetic algorithm is used to select revision and design features that lead to bugs, and five supervised techniques are compared to predict how bugs are distributed in the different modules of the (untested) design.  This information is used to allocate testing resource and target constrained random testing to target expected bugs.

\section{The Use of Machine Learning for Fault Detection}\label{sec:faultdetection}
\begin{table}[h]
    \centering
    \begin{tabular}{|m{10em}|m{8em}|m{12em}|}
        \hline
        \textbf{Type} & \textbf{Sub-Type} & \textbf{References} \\
        \hline
        \multirow{2}{*}{Evolutionary Algorithm} & Genetic Algorithm & \cite{Thamarai2010}, \cite{Bernardeschi2013}  \\
        \cline{2-3}
                                                & Genetic Program & \cite{Ravotto2008}, \cite{Bernardi2008} \\
        \hline
    \end{tabular}
    \caption{Use of machine learning for fault detection.}
    \label{tab:faultdetectionpapers}
\end{table}
Research was classified as fault detection when machine learning was used to find input sequences to cause pre-defined design errors to be detected at a DUV's output.  The primary use for fault detection is to use pre-silicon simulations to find tests for in service and post-manufacture testing.  For example, in~\cite{Bernardeschi2013} a genetic algorithm is used find DUV input patterns to detect FPGA-configuration errors caused by single-upset events.   All work in this section used genetic algorithms, and in the case of~\cite{Bernardeschi2013, Thamarai2010} operated on bit-level sequences.  
Three of the four papers in this section were not found by the structured search.  We chose to include them because their approach was similar to other work in the sampled literature and demonstrated the use of machine learning at a different level of abstraction.  For example,~\cite{Ravotto2008} explores the use of machine learning using multiple coverage metrics at different levels of abstraction to produce better coverage overall.   The work in this section also shows the use of genetic algorithms to evolve tests that hit multiple objectives~\cite{Thamarai2010, Bernardeschi2013, Bernardi2008}, which has applications in coverage closure.  There are established tools to exhaustively generate bit-level tests through formal or analytical techniques.  These tools are conventionally referred to as Automatic Test Pattern Generators.   The material in this section suggests the same ML techniques used for coverage closure also have applications at other stages in the verification process.

\section{The Use of Machine Learning For Test Set Optimisation}
\label{sec:testsetoptimisation}
\begin{table}[h]
    \centering
    \begin{tabular}{|m{10em}|m{8em}|m{12em}|}
        \hline
        \textbf{Type} &\textbf{ Sub-Type} & \textbf{References} \\
        \hline
        \multirow{2}{*}{Supervised} & Decision Trees &  \cite{Phogtat2024} \\
        \cline{2-3}
                                    & Ensemble & \cite{Parthasarathy2022a}\\
        \hline
        Evolutionary Algorithm  & Genetic Algorithm & \cite{Guo2011}, \cite{Zachariaova2016} \\
        \hline
        Unsupervised & - &  \cite{Ikram2017}, \cite{Jang2022}\\
        \hline
    \end{tabular}
    \caption{Use of machine learning for test set optimisation.}
    \label{tab:testsetoptimisationpapers}
\end{table}
Test set optimisation is similar to the test selection activity seen in coverage closure, except the machine learning operates on sets with coverage data instead of singular tests.  The objectives for the machine learning can be more diverse than seen in coverage closure.  For instance, finding the set of tests that hit all coverage points in the minimum number of CPU cycles, where unlike coverage closure, hitting a coverage point once can be regarded as sufficient~\cite{Zachariaova2016}.  The machine learning in this section can also learn from a wider range of information including design change history and previous test results~\cite{Parthasarathy2022a, Jang2022, Ikram2017}.  In particular, \cite{Parthasarathy2022a} uses a ML pipeline to predict the failure probability of an existing test and create a test set based on changes in RTL code.  The technique is notable for its use of an ensemble approach that combines the predictions of multiple (supervised) machine learning models.  Unsupervised learning techniques are used to cluster tests in~\cite{Ikram2017}, and this can be combined with Principle Component Analysis to reduce the dimensions of the learning problem~\cite{Jang2022}.

\section{Evaluation of Machine Learning in Dynamic Verification}\label{sec:evaluation}
Evaluating the performance of a proposed application of machine learning forms a crucial part of the reviewed research material.  The section summarises the designs (DUVs) and metrics authors use to evaluate their proposed techniques.

\subsection{Designs, Test Suites and Benchmarks}
\begin{figure}
    \centering
    \includegraphics[width=1.0\linewidth]{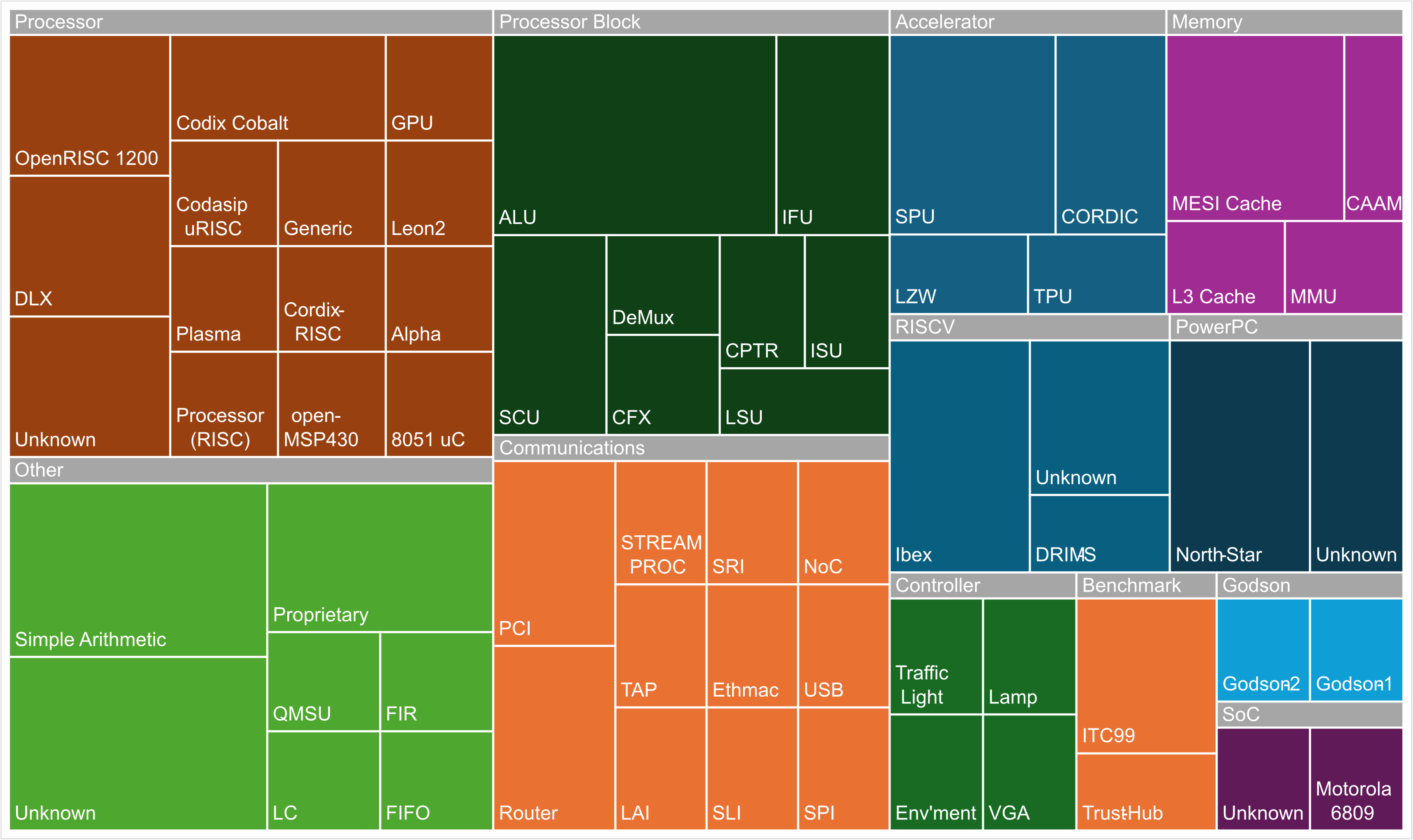}
    \caption{Designs used to test ML applications for verification.  The size of a box reflects the number of papers which use the design. ALU (arithmetic logic unit), CAAM (Cache Access Arbitration Mechanism), CFX (Complex Fixed Point), CORDIC (coordinate rotation digital computer), CPTR (Comparator), DeMux (Demultiplexer), Ethmac (EthernetMAC), FIFO (First In First Out), FIR (Finite Impulse Response filter, GPU (Graphical Processing Unit), IFU (instruction fetch unit), ISU (Instruction Sequencing Unit), ITC99 (a design from the ITC99 benchmarks), LAI (Look Aside Interface), LC (Lissajous Corrector), LSU (Load Store Unit), LZW (LZW Compression Encoder), MMU (Memory Management Unit), NoC (Network-on-Chip), PCI (Peripheral Component Interconnect includes the Express variant), QMSU (Queue Management and Submission Unit), SCU (Storage Controller Unit), Simple Arithmetic (examples include atan2, squarer and multiplier), SLI (Serial Line Interface), SPI (Serial Peripheral Interface), SPU (Signal Processing Unit), SRI (Shared Resource Interconnection), STREAMPROC (sub-block of Bluetooth protocol adapter), TAP (JTAG Test Access Port), TPU (Tensor Processing Unit), Trust-Hub (a design from the trust-hub benchmarks), VGA (Video Graphics Array).}
    \label{fig:applications_treemap}
\end{figure}

A variety of designs have been used to evaluate machine learning techniques for electronic hardware verification.  These designs range in functional complexity from simple blocks, such as ALU and comparators, to highly complex processors and system-on-chip devices~(Figure~\ref{fig:applications_treemap}).  
The range of applications shows the capability of ML to enhance the verification of different designs and at different levels of design complexity.  However, this variety makes comparing research results problematic.  It cannot be assumed an ML technique that performs well on one architecture would perform well on another at the same level of complexity or scale to different complexities. For example, it's uncertain whether the use of a genetic algorithm to verify a RISC-V Ibex core~\cite{Xia2024} would perform equally well verifying a PowerPC core or give similar results verifying a Load Store Unit.  

The challenge of comparing ML techniques is experienced across the machine learning field, leading to the creation of standard benchmarks, environments, and algorithms.  Some of these were seen in the surveyed research including supervised techniques from Python's SciKit-learn\footnote{SciKit-Learn, \href{https://scikit-learn.org/}{https://scikit-learn.org/}}.  Open-source device designs and benchmarks have also been used to evaluate the performance of EDA techniques, but their use is not universal (Table~\ref{tab:test_platforms}).  

Approximately a quarter of designs were freely accessible or described in sufficient detail to replicate easily.  The remaining three quarters included designs that an expert may be able to approximate but not reproduce exactly, such as designs to carry out simple arithmetic or implement known standards such as Serial Peripheral Interface.  Only approximately 4\% of designs were obfuscated such that the complexity of the device and its operation could not be determined. 
A small number of papers use example devices from tutorials, but these are not at the complexity level of industrial designs.  Additionally, even when open-source designs are used, including RISC-V, there remains a risk that design revisions result in the version used in a piece of research being unavailable or unknown.

This lack of standardisation may delay the progress and adoption of machine learning for coverage closure relative to other areas.  Research on coverage closure is frequently conducted in collaboration with private companies, where the pursuit of commercial advantage often restricts the availability of designs alongside the research findings.  One approach taken by~\cite{Parthasarathy2022a, Huang2022, Nazi2022} which balances the needs for IP protection with open research is to include results from an open source design alongside those from proprietary devices.

\begin{table}[h]
    \centering
    \begin{tabular}{| m{15em} | m{10em} | }
         \hline
         \textbf{Design Repositories} &  \textbf{Used in}\\
         \hline
         ITC'99 & \cite{Bernardeschi2013, Yu2002}  \\
         \hline
         Trusthub & \cite{Mondol2024}  \\
         \hline
         Opencores & \cite{Shen2005, Chang2010, Guo2014, Gadde2024, Bhargav2021, Ghany2021}   \\
         \hline
         \hline
         \textbf{Processors} &  \\
         \hline
         RISC-V Ibex & \cite{Parthasarathy2022a, Huang2022, Xia2024} \\
         \hline
         OpenSPARC & \cite{Guzey2008}  \\   
         \hline
         DRIM-S & \cite{Guzey2008}  \\
         \hline
         Leon2 & \cite{Corno2004b}  \\
         \hline
         \hline
         \textbf{Tools} & \\
         \hline
         CoCoTb Python package & \cite{Tweehuysen2023}  \\
         \hline
         RISC-DV & \cite{Huang2022}  \\   
         \hline
    \end{tabular}
    \caption{Open source platforms used for evaluating machine learning for dynamic verification.}
    \label{tab:test_platforms}
\end{table}

\subsection{Measuring Performance}

\subsubsection{Metrics}
Metrics are used by authors to measure the performance of a machine-learning application.  In the sampled literature, six categories of metrics were identified.  A description of each is given in Table~\ref{tab:metricgroups}.

\begin{table}[h]
    \centering
    \begin{tabular}{|p{0.17\textwidth}|p{0.6\textwidth}|}
        \hline
        \textbf{Group Name} & \textbf{Description} \\ 
        \hline
        \textbf{Learning \phantom{xxx} Performance} & Classical ML and statistical metrics that measure how well the ML fits the application. Metrics include: Measure Square Error~\cite{Ismail2021a}, F and F2 score, recall, accuracy, precision, loss learning rate, number of correct predictions, and false positives.   \\
        \hline
        \textbf{Application \phantom{xxx} Performance} & Metrics common in applications related to coverage closure. The most common measure is coverage as a percentage. Other values include hit rate~\cite{Hsueh2006}, the number of coverage points hit~\cite{Liang2023} and test diversity~\cite{Romero2009}.   \\
        \hline
        \textbf{Stimulus Count} & Used to measure the test resource required. Examples include the number of times the ML updates constraints, the number of instructions or transactions simulated, the number of simulations, and the number of tests.   \\
        \hline
        \textbf{Execution Time} & An alternative to counts for measuring test resources. Authors use terms including simulation time, execution time and wall time.   \\
        \hline
        \textbf{ML Overhead} & Measure the additional resources a machine learning method adds to verification. Some research measures this extra cost as total overhead time, others use more granular measures, including the time to train a model, the prediction time, and the time spent generating test patterns that are discarded.   \\
        \hline
        \textbf{Other} & Used for specialist applications, including the number of sampled modules~\cite{Roy2018} and metrics used by a commercial tool~\cite{Ikram2017}   \\
        \hline
    \end{tabular}
    \caption{Metrics used to assess the performance of machine learning methods in dynamic microelectronic verification.}
    \label{tab:metricgroups}
\end{table}

\begin{figure}[h]
    \centering
    \includegraphics[width=0.7\linewidth]{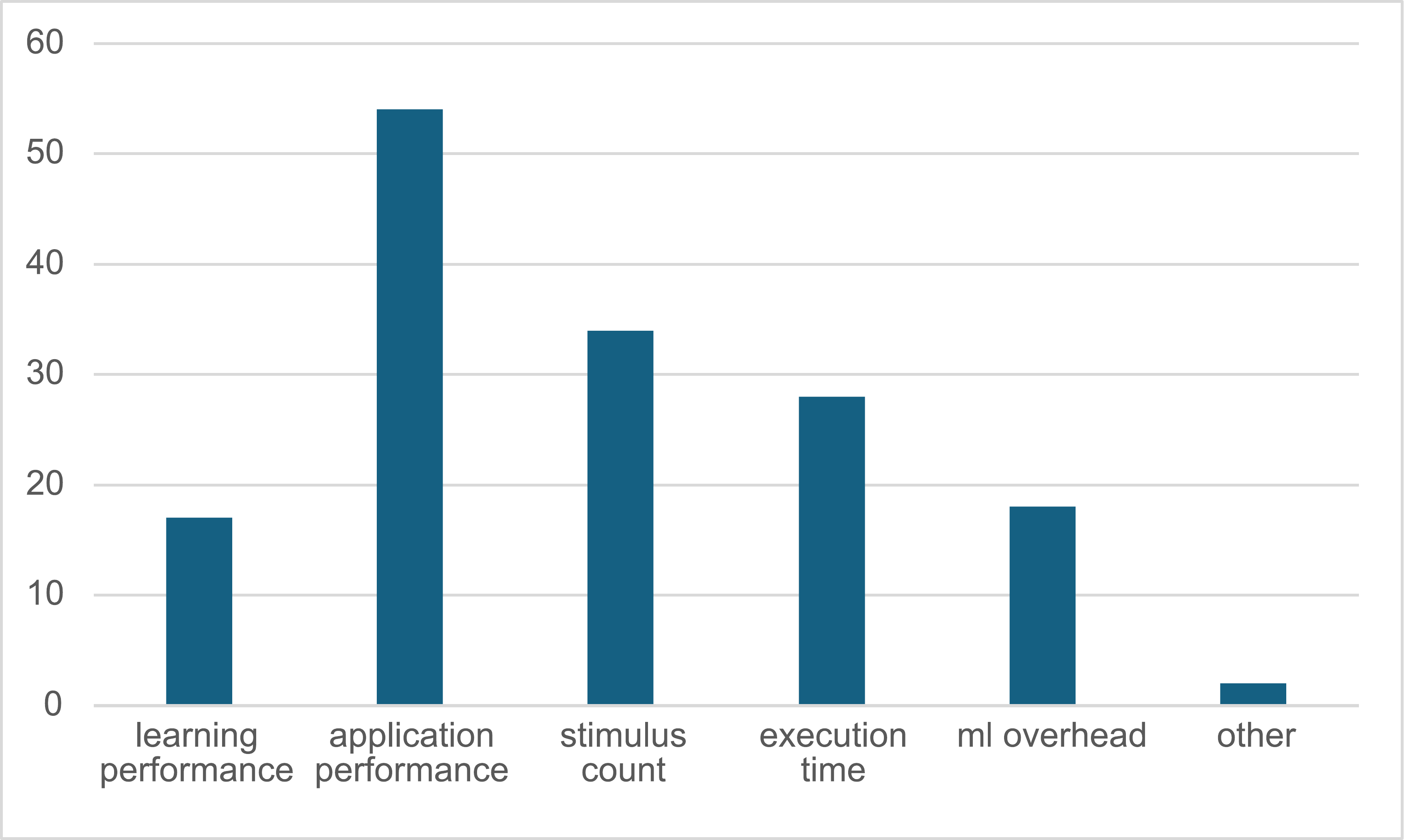}
    \caption{A count of the type of metrics used to assess machine learning for microelectronic design verification. Metrics of the same type are not double-counted within the same piece of material.  If a single piece of research material employs more than one metric of the same type, it only increases the count of that metric type by one.  Measures relating to task performance were used most frequently.}
    \label{fig:countofmlmetrics}
\end{figure}

Application performance emerged as the most widely used metric for assessing techniques. In contrast, learning performance and ML overhead were less commonly reported than one might expect in applied machine learning research (Figure~\ref{fig:countofmlmetrics}). An argument is that application performance reflects the real-world benefits of using a technique. However, classic metrics for learning performance provide insights into an algorithm's 'fit' to the data and environment. Every learning technique incurs an associated resource cost, making it crucial to understand the cost-to-performance benefit when comparing techniques. For industry practitioners looking to adopt a technique, the tendency of research to report only the benefits hinders meaningful comparison.

\subsubsection{Baselines}
\begin{figure}[h]
    \centering
    \includegraphics[width=0.7\linewidth]{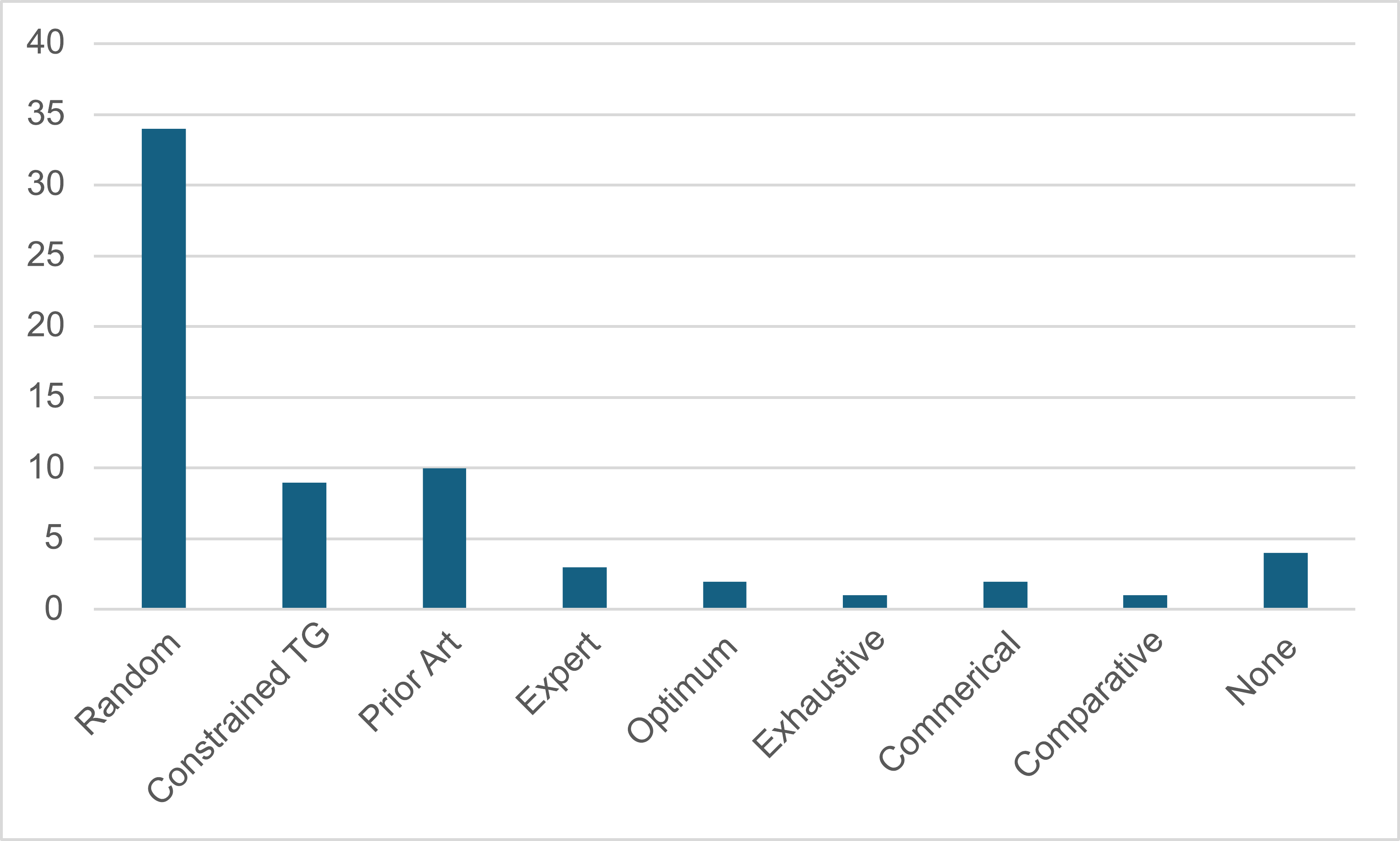}
    \caption{A count of the baselines seen in the literature for assessing the performance of a machine learning application for microprocessor verification.}
    \label{fig:baselines_bar}
\end{figure}
Measures of performance, particularly those relating to resources used, are often compared to a baseline.  The most commonly used baseline is random-based methods (Figure~\ref{fig:baselines_bar}).  These methods include randomising instructions, constraints or pre-generated tests depending on the specific use and application of machine learning.  Research that proposes more than one method or evaluates a family of ML methods made comparisons between the techniques~\cite{Mondol2024, Dinu2021}.  A small number of applications used either expert-derived parameters, optimum results, or the ground-truth design as a baseline that an ideal machine learning application could achieve. 
 
Using random-based methods as a baseline is advantageous because these methods are the most commonly used in industry and supported by existing simulation-based workflows. Random also acts as a ``lowest common denominator'' to circumvent the time and complexity of replicating ML methods proposed by other authors.  Other sections of this review highlight the lack of openly available information, data sets and designs. 
 In the absence of being able to replicate work, random-based methods are a means to compare performance between different applications of ML.  However, caution is needed because performance vs random does not measure how well a technique generalises.  The comparative studies demonstrate that different ML methods perform differently for the same application.  Therefore, a method that performs well against random in one application may not perform well in another.  This makes the insight gained from research that compares ML methods valuable.

\section{Challenges and Opportunities} \label{sec:challengesopportunities}
The surveyed material presents a rich and varied set of machine learning techniques and applications for verifying electronic designs.   The number of publications on this topic has increased and showcases successes for EDA practitioners to use or build upon.  However, trends were seen that hinder progress:
\begin{itemize}
    \item[--] A lack of standard benchmarks, withholding code and data, and obfuscating work undertaken with private companies make it difficult to replicate results and measure progress.
    \item[--] Techniques are evaluated on simple designs without comparisons to other well-established and effective methods other than random.
    \item[--] Research rarely explores whether a technique will generalise beyond the application tested or scale to real-world systems.
    \item[--] It is rare to see work justify the choice of machine learning technique and how it is applied.  
    \item[--] Research is confined to a tool or ML type, and it is rare to see an exploration of alternative methods.  If comparisons between techniques are made, these tend to be within the same family of techniques.
    \item[--] The criteria for assessing the success of a technique are confined to a single metric and do not capture the criteria for real-world adoption.
    \item[--] Research treats verification as a one-shot problem, whereas in industry it is a rolling process throughout development.
\end{itemize}

These trends create problems of generalisation, replication and assessment.  This section discusses the challenges these trends create and the opportunities for progress.

\subsection{Existing Industry Practice}
A tenancy was seen for research to treat EDA verification as an academic problem in which the performance of a particular technique is the only measure of success.  In real-world use, EDA verification is a tried and tested industrial process.  The challenge for research is to account for this incumbent process and the ease by which a technique can be implemented.  The qualities in Section~\ref{sec:testbenchqualities} highlight a range of criteria, which is one step towards appraising techniques in the context of real-world use.  Research that provides interfaces between learning methods and existing test bench designs and generalises between verification environments is also valuable for real-world adoption.

Dynamic-based verification of electronic hardware creates a large amount of labelled data.  This data is generated over time on a design experiencing frequent incremental changes.  Changing ground-truth relationships caused by these design revisions and the availability of new data create opportunities for research on machine-learning techniques designed for dynamic environments.  Research was seen that used classical analysis and statistics in this design environment.  For measuring the difference between two versions of a design to inform testing and classical statics to exploit the volume of data generated by typical verification progress.  However, there are research opportunities that use machine learning with the design changes and large volumes of data seen in industrial development, particularly techniques that are useable at the start of a project and improve over time, such as hybrid techniques.

\subsection{Similarities with Test-Based Software Verification}
Testing software and hardware designs are fundamentally similar tasks; both disciplines aim to establish the correct operation of a function relative to a specification by applying inputs and monitoring the output.  However, it is rare to find research that translates between the software and hardware testing domains.  Despite the two domains appearing to operate in isolation, many of the trends identified were also identified in a recent survey for machine learning in software testing~\cite{Fontes2023}.  Specifically, overuse of simple examples, lack of standardised evaluation criteria, unavailable code and data, and research that does not investigate whether techniques will scale to real-world systems, justify the choice of technique or compare alternatives.  Given the similarities in the domains, there is an opportunity to coordinate research efforts.  

An example of a technique translated from software to hardware testing is fuzzing~\cite{Canakci2021}.  It has been researched for verifying applications of RTL on FPGAs~\cite{Laeufer2018}, and implemented based on existing tools used in software testing~\cite{Ruep2022}.  Fuzzing is a technique that was first proposed for software testing and has seen real-world adoption by leading companies including Microsoft\footnote{Microsoft, ``microsoft/onefuzz'', \url{https://github.com/microsoft/onefuzz}} and Google\footnote{Google, ``google/clusterfuzz'', \url{https://github.com/google/clusterfuzz}}.  The method has similarities to constrained random and GA approaches, that were a subject of research and use in hardware verification before fuzzing was proposed.  Current research does not directly compare fuzzing with constrained random and ML techniques, so it is unknown if it is more efficient for hitting hard-to-hit points.  However, the advantages of fuzzing are the low setup cost, simple operation and improving performance over time.

Research in software testing not only introduces new techniques but also offers EDA practitioners valuable insights into methods less prevalent in the hardware domain. For example, while reinforcement learning has been extensively explored for testing sequentially driven software, particularly GUIs~\cite{Fontes2023}, its application in micro-electronic verification remains limited to basic problems. The software domain could also inspire innovative uses of machine learning in hardware verification. This review highlights that machine learning applications in hardware verification are predominantly focused on coverage-related use cases (Section~\ref{sec:usecases}). In contrast, a recent review of ML in software testing revealed a similar focus on coverage but also identified more material on enhancing the effectiveness and efficiency of existing methods than is currently seen in the hardware domain~\cite[Section 4.3]{Fontes2023}

Overall, greater coordination between research in software and hardware testing presents opportunities for knowledge transfer and synthesis. This can increase the number of applications and advance the use of machine learning for dynamic-based verification.

\subsection{Evaluating the Strengths and Weaknesses of ML Techniques}
The only example seen of research that compared two different types of ML techniques was in~\cite{Pfeifer2020}, where a reinforcement learning (RL) technique was compared to an existing genetic algorithm.  No research was found comparing supervised techniques with RL (or Evolutionary Algorithm (EA)) methods.  This gap presents an opportunity for future research to examine the relative strengths of different types of ML techniques, particularly for coverage closure in relation to their use of training data.

Supervised methods trained offline often used data acquired through other means, such as random stimulus~\cite{Wang2018, Katz2011, Ghany2021}.  Additionally, it has been shown that random stimulus outperformed RL for low coverage percentages, negating its benefit over supervised methods at the start of learning.  The open question is whether RL or supervised techniques are more efficient overall at reaching the hard-to-hit coverage points.  Specifically, does the greater control an RL or EA method have to explore the space at the start of learning enable it to reach coverage closure with fewer simulations, or is the often randomly created dataset for supervised methods, which learns offline and cannot influence their own training data, just as good?  To address these questions, it is recommended that future research:
\begin{itemize}
    \item[-] \textbf{Conduct comparative studies:} Perform direct comparisons between supervised, RL, and EA methods across various benchmarks to identify their strengths and weaknesses in different scenarios.
    \item[-] \textbf{Analyse training data utilisation:} Investigate how the source and quality of training data impact the performance of each ML technique, particularly in achieving coverage closure.
    \item[-] \textbf{Evaluate efficiency:} Measure the efficiency of each technique in terms of the number of simulations required to reach high coverage, considering both initial learning phases and long-term performance.
    \item[-] \textbf{Explore hybrid approaches:} Examine the potential benefits of combining supervised and RL/EA methods to leverage the strengths of both approaches
\end{itemize}

\subsection{Use of Open Source Designs and Datasets}
The range of applications, benchmarks, and metrics used to assess ML techniques can makes it challenging to compare techniques (Section~\ref{sec:evaluation}).  Also, those wishing to apply a technique in a different application would be unable to easily establish the differences between the tested environment and their own. Greater use of open source designs and production of common data sets are potential solutions.

Benchmarking machine learning verification techniques on open source designs enables others to replicate the work and compare the performance of techniques.  Some of the surveyed works already use open source designs.  To enable meaningful benchmarks, open source coverage models, verification environments and standardised test procedures are also needed.  Taking inspiration from the wider field of machine learning, a similar need for standardised testing environments led to the development of OpenAI Gym~\cite{Brockman2016} in the reinforcement learning community.

Data is central to most machine learning techniques.  One of the present difficulties in hardware verification is that acquiring data requires expertise in running test benches.  This is a specialist skill that includes knowledge of SystemVerilog, scoreboards, monitors, and coverage definition; skills not necessarily possessed by machine learning experts.  Again, taking inspiration from the wider machine learning community datasets such as ImageNet~\cite{Deng2009} provided the platform for significant breakthroughs in the use of machine learning for image classification\footnote{Ksenia Se, ``The Recipe for an AI Revolution: How ImageNet, AlexNet and GPUs Changed AI Forever'', \url{https://www.turingpost.com/p/cvhistory6}}.  The need for large-scale, open datasets was also one of the recommendations of a recent survey into the use of machine learning from a verification industry perspective~\cite{Yu2023}.  

Open source designs, including RISC-V\footnote{\url{https://riscv.org/}}, have matured to the point where they are used in commercial products and openly supported by companies including Thales\footnote{\url{https://www.thalesgroup.com/en/group/journalist/press-release/thales-joins-risc-v-foundation-help-secure-open-source}} and Western Digital\footnote{\url{https://blog.westerndigital.com/risc-v-swerv-core-open-source/}}.  There is an opportunity for commercial companies to produce datasets, benchmark environments and metrics for these open source designs and challenge the machine learning community to find high performing, commercially viable, machine learning techniques to verify them.  This would enable industry to drive research in a direction that is relevant and commercially beneficial.

\subsection{The Prevalence of Open Source Designs in Commerical Products}
The increasing maturity of open-source designs of processor cores raises the possibility of their use by electronic design companies unaccustomed to the verification needs of core design. 
 For reference, ARM cores are subject to many hours of simulation-based testing running on high-performance clusters.  A typical company using an open-source design does not possess the computational resources, expertise, or access to the EDA tools required to achieve similar levels of verification.  Therefore, a need and opportunity exist for open research that can be used by small electronic design houses to verify their applications based on open-source core designs.

\section{Challenges for Future Research}
The results of this review highlight the difficulties of applying machine learning to the verification of microelectronic devices in a real-world project.  There are many examples of successful applications of machine learning, but also many configurations of elements that affect the learning.  These elements include abstraction level of both the input and output spaces of the ML model, what the machine learning controls, whether the ML is used to target a single coverage hole or many holes, the hyper-parameters of the ML models, and more.  What this review concludes is that while there are many successful applications of ML for verification, there is very little understanding of why the application was successful.  This information is crucial to generalise a technique to different applications.

To gain widespread adoption, the use of machine learning techniques for verification could look to the adoption of formal techniques as a case study.  Once seen as requiring complex setup and specialist skills, formal techniques are now more accessible to verification engineers.  This has been achieved by offering guided workflows to configure and run the tool as a ``push button'' operation in industrial EDA software suites. 

In summary, the questions for future research into the use of ML for verification are as follows.
\begin{itemize}[nosep]
    \item[-] Why does a machine learning technique work for a specific application?
    \item[-] How would the technique transfer between different applications?
    \item[-] What are the limitations of the technique?
    \item[-] What domain knowledge, assumptions, and constraints are needed to apply the technique?
\end{itemize}

\section{Acknowledgments}
The authors acknowledge the assistance of Maryam Ghaffari Saadat in the preparation of this review.  

\bibliography{menderly_lib20250302}

\end{document}